\documentclass[10pt,journal,compsoc]{IEEEtran}
\usepackage{amsmath}
\usepackage{amssymb}
 \usepackage{amsthm}
 \usepackage{color}
 \usepackage{tikz} \usetikzlibrary{fit,positioning}
 \usepackage{graphics}
 \usetikzlibrary{arrows}
 \usepackage{hyperref}
 \usepackage{subfigure}
 \usepackage{tikz}
 \usetikzlibrary{arrows,positioning,shadows}
 \usepackage{standalone}
 \usepackage{algorithmic}
 \usepackage{algorithm}
 \usepackage{tabularx,ragged2e,booktabs,caption}

\begin{document}

\title{A simple multiforce layout for multiplex networks}

\author{Zahra~Fatemi,
        Mostafa~Salehi*,
        and~Matteo~Magnani
\IEEEcompsocitemizethanks{\IEEEcompsocthanksitem Zahra Fatemi and Mostafa Salehi (*corresponding author) are with University of Tehran.

E-mail: fatemi.z@ut.ac.ir, mostafa\underline{\space}salehi@ut.ac.ir
\IEEEcompsocthanksitem Matteo Magnani is with Uppsala University.

Email: matteo.magnani@it.uu.se}
}

\IEEEtitleabstractindextext{%
\begin{abstract}
We introduce \emph{multiforce}, a force-directed layout for multiplex networks, where the nodes of the network are organized into multiple layers and both in-layer and inter-layer relationships among nodes are used to compute node coordinates. The proposed approach generalizes existing work, providing a range of intermediate layouts in-between the ones produced by known methods. Our experiments on real data show that multiforce can keep nodes well aligned across different layers without significantly affecting their internal layouts when the layers have similar or compatible topologies. As a consequence, multiforce enriches the benefits of force-directed layouts by also supporting the identification of topological correspondences between layers.
\end{abstract}

}

\maketitle
\IEEEdisplaynontitleabstractindextext

\IEEEpeerreviewmaketitle

\section{Introduction}
\label{sec:Intro}
\IEEEPARstart{N}{etwork} analysis is a widely applied discipline, due to the fact that many realities can be modeled as sets of interconnected entities -- for example, social networks, technology networks, transportation networks, utility networks and biological networks. In the following, we use the term \textit{monoplex} to indicate a network consisting of nodes with a single type of relationship among them. The simplicity of monoplex network models is at the same time a strength, making them applicable to diverse contexts, and a limit, because it may hide many details of the modeled reality. Therefore, multiplex networks are often used as a richer but still simple and general model, as they allow multiple types of relationships between nodes. For example, two people in an online social network like Facebook can be friends, while being colleagues in a work environment.

Many aspects of multiplex networks from spreading processes to structural measures such as clustering coefficient and node centrality have been recently investigated \cite{DBLP:conf/asonam/MagnaniR11, Berlingerio2012, Sole-Ribalta:2014:CRM:2615569.2615687, battiston2014structural, Kivela2014, salehi2015spreading, Dickison2016}, building on a long-standing literature in Social Network Analysis \cite{Wasserman1994}, but only a few works have specifically investigated how to draw multiplex networks.
Several layouts have been proposed to visualize monoplex networks, arranging nodes so that users can easily identify special network structures like hubs or communities, and quickly locate important nodes \cite{Landesberger2011}, but drawing multiplex networks is significantly more challenging.

\begin{figure}[tp]
  \begin{center}
    \subfigure[Flattened visualization]{\label{multivisua1}\includegraphics[scale=0.25]{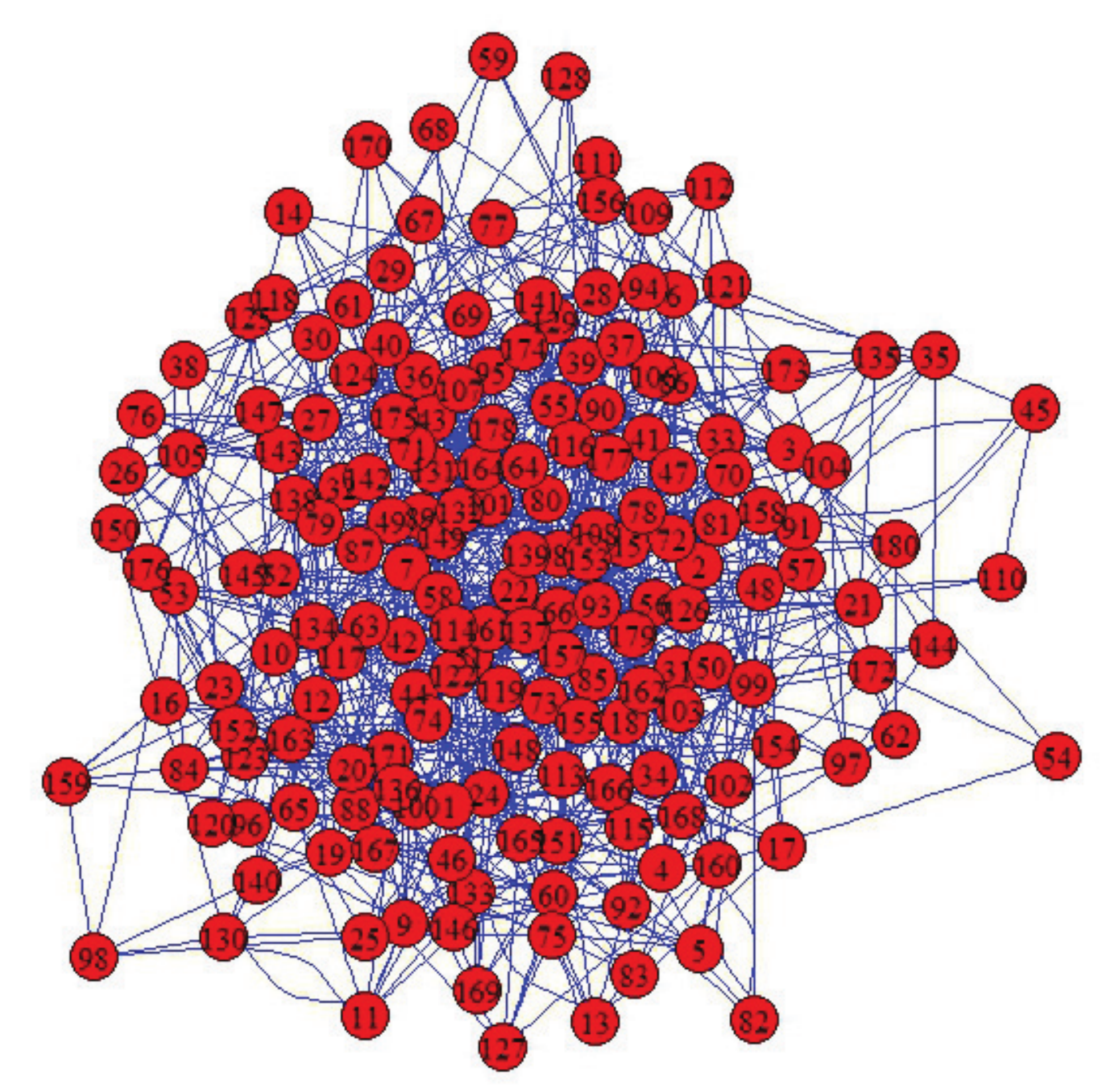}}
    \subfigure[Sliced visualization]{\label{multivisua2}\includegraphics[scale=0.25]{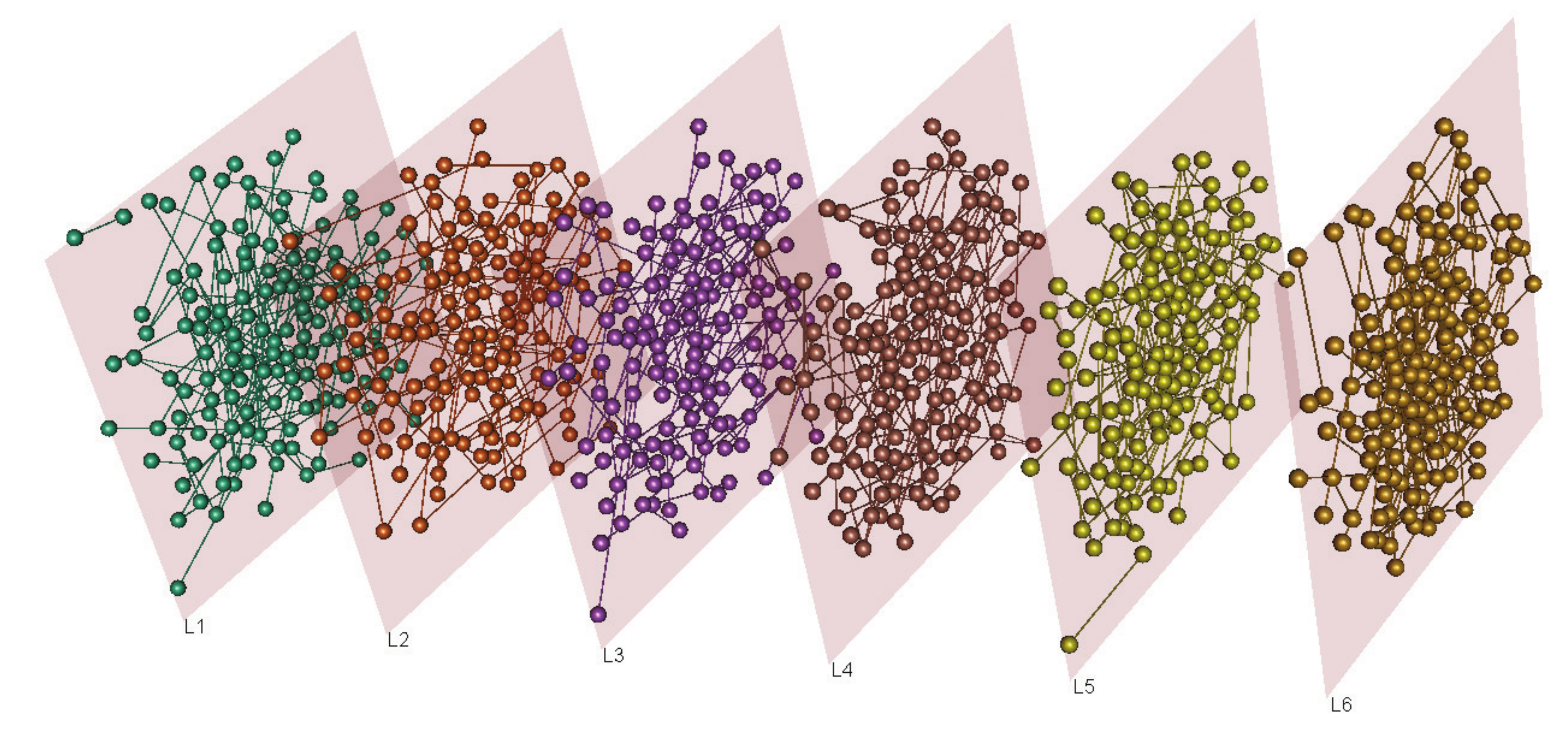}}
    \\  
  \end{center}
  \caption{
Two alternative ways of visualizing a multiplex network with 180 nodes and 863 edges of 6 types
  } 
  \label{multivisua}
\end{figure}

A multiplex network visualization should support two main types of tasks: the analysis of the single layers, each corresponding to a traditional monoplex network for which existing graph layouts can be used, and the analysis of the relationships between layers, for which it is useful to identify where the same nodes or network structures are located on the different layers \cite{Dickison2016}.
Representing multiple relationship types in the same graph as in Figure~\ref{multivisua1} 
can quickly lead to a very dense representation hiding relevant network structures \cite{2015arXiv150101666M}. Therefore, different relationship types can be organized into different layers, with the same node replicated on multiple layers, as in Figure~\ref{multivisua2}. This approach has been used in the literature to represent different types of multiplex networks, from traffic networks \cite{Kurant2006,de2014muxviz} to biological \cite{de2014muxviz} and social/historical \cite{Padgett2006,DBLP:conf/asonam/MagnaniR11} networks.

In the literature, two main approaches for visualizing multiplex networks sliced into layers have been used: keeping the same layout in all layers, meaning that each node keeps the same position in all layers and all its replicas would result aligned on a straight line if the layers are visualized one besides the other, or visualizing each layer independently of the others. The first approach focuses only (or mainly) on inter-layer relationships, highlighting node correspondences across layers, while the second focuses only on in-layer edges. In our view these approaches are just two specific cases of a more general method. To fill the gap between these two extremes we propose to define a range of intermediate layouts with a controllable balance between in-layer and inter-layer relationships.
We call our general layout \textit{multiforce}.

\begin{figure}[tp]
\begin{center}
\includegraphics[scale=0.3]{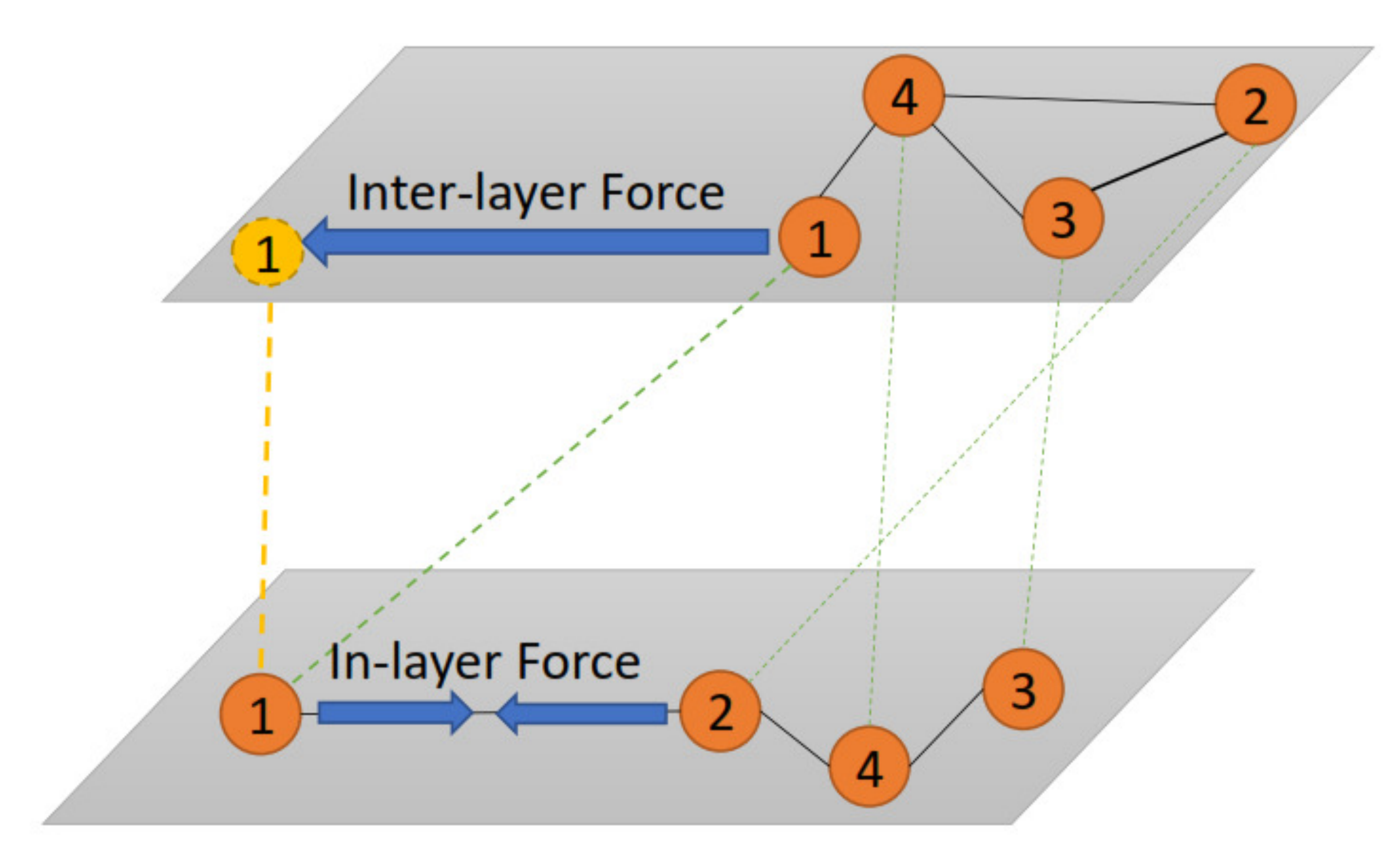}
\end{center}
\caption {The effect of in-layer and inter-layer forces on node positions}
\label{twolayer}
\end{figure}

Multiforce is based on a force-directed algorithm (Fruchterman-Reingold) and uses
two main types of forces: \textit{in-layer} and \textit{inter-layer}, that can be tuned to impact specific layers more or less than others. In-layer forces attract neighbors inside the same layer, making them closer, as in traditional layouts for monoplex networks. Inter-layer forces try to align instances of the same node on different layers\footnote{In theory inter-layer forces can also be used to visualize more general networks, where edges can cross layers, but in this work we focus on multiplex networks.}. Figure~\ref{twolayer} gives an intuition of how these forces operate. In addition, we also use repulsive forces as in the original algorithm.

A specific problem of multiplex network visualization is that different layers can expose different structures, for example two nodes connected on one layer may be disconnected on another. Therefore, in general it is impossible to produce a good layout for each layer and keep nodes perfectly aligned across different layers at the same time. As an example, node 2 in the lower layer of Figure~\ref{twolayer} would be attracted towards node 1 on the same layer because of the edge between them, but this would increase the distance between the position of node 2 on the lower layer and the position of node 2 on the upper layer. In this way we would negatively affect our visual understanding of the relationships between the different layers, losing our ability to quickly locate the same node across layers.

To enable an objective evaluation of our approach with respect to its ability to keep nodes aligned across layers but also preserve layer-specific structures we have defined two quality metrics representing these two criteria, called respectively \emph{external fit} and \emph{internal fit}. These are naturally defined as the forces which would be active on the nodes if we were only trying to respectively align them across layers or draw them according to the internal structure of the layer. We have then performed an evaluation on several real networks showing how existing approaches would optimize only one of these two metrics, while our algorithm can obtain good scores on both at the same time. In addition to this evaluation, we have also executed our algorithm on a simple dataset to characterize the resulting diagrams. A practically valuable result is that a visualization of two layers (or relationship types) whose structures are similar only at specific locations would highlight these similarities, by keeping only those nodes well-aligned --- something that global layer similarity measures cannot capture.

In Section \ref{sec:Related Work} we quickly review existing layouts for monoplex and multiplex networks. In Section \ref{propoesedlayout} our multiforce algorithm is introduced. Section \ref{sec:char} presents an example using a small synthetic dataset, to characterize the impact of different settings, while Section \ref{sec:eval} proposes an evaluation on real data.

\section{Related Work}
\label{sec:Related Work}
In this section we describe previous works on monoplex and multiplex network visualization.

\subsection{Monoplex Network visualization}
Many layouts have been designed to visualize monoplex networks. Here we briefly review the ones that are more relevant for our approach. For an extensive review, the reader may consult \cite{Landesberger2011}.

\textit{Multi-scale layouts} first create some core sub graphs, then they add other nodes until all nodes and edges are added \cite{Landesberger2011, 1173159}. \textit{Random layouts}  \cite{Diaz:2002:SGL:568522.568523} and \textit{circular layouts} \cite{some} are two categories of layouts which are appropriate for small graphs with few nodes and edges, because they do not consider aesthetic criteria: many edge crossings and node overlappings can appear. 

Among the most used visualization methods, \textit{force-directed} algorithms consider a graph as a physical system where forces change the position of nodes. The two best known force-directed layouts are \emph{Fruchterman-Reingold} \cite{Fruchterman:1991:GDF:137556.137557} and \textit{Kamada-Kawai} \cite{Letters1989, Moody2005}. In the Fruchterman-Reingold layout nodes have repulsive power and push other nodes away, while edges attract neighbor nodes. In this layout nodes are considered as steel rings haing similar loads and edges are like springs attracting neighbouring rings. This algorithm consists of three main steps. First, all nodes are distributed randomly. Second, repulsive forces separate all nodes. The value of repulsive force depends on the positions of the nodes. Third, for each edge and based on the position of nodes after repulsion, attractive forces are calculated \cite{Fruchterman:1991:GDF:137556.137557}.
In the Kamada-Kawai layout an energy function is defined for the whole graph based on shortest paths between nodes, and positions are iteratively updated until the graph's energy is minimized \cite{Letters1989}. 

Bannister et al.~\cite{gravity} proposed a force-directed layout to change the position of nodes so that more graph-theoretically central nodes pushed towards the centre of the diagram. In this algorithm, an additional force called gravity is used to change the position of more central nodes. For each node $v$ in a graph $G$ the position of the node is influenced by the following force:
\begin{equation}
\label{gravityforce}
I[v] =\sum_{u,v \in V}f_r(u,v)+\sum_{(u,v) \in E}f_a(u,v)+\sum_{v \in V} f_g(v)
\end{equation}
where $f_r$ and $f_a$ are respectively repulsive and attractive forces, and $f_g$ is the gravity force, measured as:
\begin{equation}
\label{gravityforce2}
f_g(v)=\gamma_t M[v](\xi - P[v])
\end{equation}
In this equation $M[v]$ is the mass of node $v$, which can be set according to the node degree, $P[v]$ is the position of $v$, $\gamma_t$ is the gravitational parameter 
and $\xi = \Sigma_v P[v]/|V|$ is the centroid of all nodes. Notice that forces in the equations above are vectors. 

Other extensions of force-directed layouts have considered the inclusion of additional domain-specific information in the definition of the forces. An example is \cite{Gan-2013}, where the defiinitions of attractive and repulsive forces include terms representing trust in social networks.
 
Force-directed methods have two main drawbacks. First, they are typically suitable for networks with at most 100 nodes, because attractive forces result in hiding some relationships among nodes and node-overlapping increases. Second, their run time is high in comparison with other approaches. However, they are very popular, because they often practically succeed in separating clusters and increasing graph readability.
To address these shortcomings, layout algorithms are sometimes split into multiple phases, with an initial preprocessing of the data to generate good starting node dispositions or to reduce data complexity. As an example, Gajer et al.'s method first partitions the graph into subgraphs \cite{GAJER20043}. The smallest subgraph is then processed independently and thus more efficiently. Afterwards, a force-directed refinement round changes the values of initial node positions and the next smallest subgraph is added to the previous one, with these steps being repeated untill all nodes have been processed. \cite{Baur2005} follows the same steps, setting initial node positions in a different way.
Similarly, Peng et al.~proposed
the Social Network Analysis Layout (SNAL) by also separating a network into subgroups, analyzing relationships between them and positioning nodes based on their relative centralities \cite{peng}.

Another family of layouts, that  can also be combined with force-directed algorithms, are \textit{constraint-based} layouts \cite{Dwyer}. These layouts force nodes to appear at specific positions. For example, nodes are placed on a frame in a way that they do not overlap, or are horizontally and vertically aligned, as in the \textit{orthogonal} layout \cite{Landesberger2011, Dwyer}. In these layouts, it is more difficult to isolate special structures like communities and time complexity is noticeably high.

One important assumption in graph drawing is that there is a correspondence between some aesthetic features of the diagrams and their readability. Therefore, some visualization algorithms explicitly target these features. One such criterion is that too many edge crossings make a graph more difficult to interpret.
The crossing number, $cr(G)$, of a graph G is the smallest number of crossings appearing in any drawing of G \cite{schaefer2013graph}. 
Several algorithms have been proposed to reduce edge crossing in monoplex networks. For example, Shabbeer et al.~\cite{shabbeer2010optimal} developed a stress majorization algorithm. In \cite{schaefer2013graph} and \cite{Buchheim_crossingsand} the concept of edge crossing is elaborated and equations for measuring the number of edge crossings in different graphs are reviewed.
Another aesthetic feature impacting graph readability is node overlapping. Two popular methods to reduce node overlapping are proposed in \cite{Huang20072821, Kumar:2009:VCD:1529282.1529685}.

\subsection{Multiplex Network Visualization}
Different methods have been discussed for visualizing multiplex networks. We can categorize these methods into three main classes: 
\begin{itemize}
\item
\textbf{Slicing:}
One way of visualizing multiplex networks is to show each layer or relationship type as a monoplex network and to connect these monoplex networks using inter-layer edges \cite{de2014muxviz}. The layers can have aligned layouts or independent layouts. Aligned layouts help users find similar nodes in different layers by forcing the same node to have the same coordinates on all layers, but structures existing only on one layer (for example communities) may not be clearly visualized. Independent layouts can show specific structures of each layer, but may hide inter-layer patterns \cite{2015arXiv150101666M}. 
\item
\textbf{Flattening:}
In these methods, all nodes and edges are placed on the same plane. In a \textit{node-colored} network, nodes from different layers are shown with different colors \cite{Kivela2014}, while for multiplex networks colors can be used to distinguish edges of different types. Apart from suffering from the same problems of aligned slicing, the disadvantage of this method is that for networks with high edge density relationships among nodes can be hidden by edges from non-relevant layers and readability quickly decreases due to the network's clutter \cite{2015arXiv150101666M}.
\item
\textbf{Indirect:}
This approach tries to visualize information derived from the network instead of directly visualizing the original layers, which are considered to be too complex to allow a simple visual representation.
Renoust et al.~\cite{10.1111:cgf.12644} proposed a system for visual analysis of group cohesion in flattened multiplex networks. This system, called \textit{Detangler}, creates a so-called substrate network from unique nodes of the multiplex network and a so-called catalyst network from edges of different types. 
Erten et al.~\cite{Erten2004} proposed three modified force-directed approaches for creating slicing, flattened and split views of multilayer networks by considering edge weights and node weights. The weights of nodes and edges are based on the number of times they appear in different layers. In this approach interlayer relationships between nodes are ignored and node weights are the same for all nodes when multiplex networks are visualized, so this approach does not consider the specific features of the networks targeted in our work.

An extreme case of indirect methods, that we mention for completeness, consists in not visualizing nodes and edges at all but only indirect network properties, such as the degree of the nodes in the different layers or other summary measures \cite{de2014muxviz,2015arXiv150101666M,Redondo2015}. These approaches are complementary to graph drawing, and can also be used in combination with our proposal.
\end{itemize}
Our method belongs to the slicing class, and is different from existing approaches because it allows a balancing of the effects of in-layer and inter-layer relationships.

\section{The Multiforce Layout}
\label{propoesedlayout}

Multiforce extends the Fruchterman-Reingold algorithm \cite{Fruchterman:1991:GDF:137556.137557}, and as mentioned in the introduction is based on two types of attractive forces. The nodes are positioned on a set of planes, one for each layer or type of relationship -- this setting is sometimes called 2.5-dimensional, because it looks 3-dimensional but the z-coordinates of the nodes are fixed and limited to the number of planes/layers. In-layer forces, that can be weighted differently in each layer, attract pairs of nodes connected on the same layer. Inter-layer forces influence the position of nodes in different layers connected by inter-layer edges, or corresponding to the same node in the case of multiplex networks. 

The pseudo-code of multiforce is presented in Algorithm \ref{mylayout}.
The algorithm takes a multiplex network $G = (N, L, V, E)$ as input, where $N$ is a set of nodes, $L$ a set of layers, $(V,E)$ is a graph and the elements of $V$ are pairs $\langle$node, layer$\rangle$. We notate $v$.layer the layer of an element $v \in V$ and $v$.node the node corresponding to an element $v \in V$.

Lines 13-29 are the same as in the original algorithm, and compute the displacement of each node based on its neighbors (attractive forces) and other nodes (repulsive forces), with the addition of weights that can be used to specify on which layers the layout should be computed according to the original algorithm (27-28).  Lines 30-37 extend the original algorithm and compute the displacement caused by the position of the node on other layers, to control node alignment. This is also weighted, to allow the user to turn this feature on and off for all or some layers (34-35).

Some details of the algorithm can be changed without affecting its underlying idea. First, we can modify lines 6-12 to assign the same initial random coordinates to the same node across different layers, anticipating line 8 before the for loop. A weighting factor $\textrm{INLA}[v]$ can also be added at line 20, so that both attractive and repulsive forces are reduced or reinforced together --- in practice, this does not seem to have a significant effect on the result; all diagrams in Section \ref{sec:char} have been computed without these weights. Finally, lines 41 and 42 have been retained from the original algorithm and ensure that the nodes do not exit the frame specified by the user, but are not necessary if the final coordinates are re-scaled to fit it.

\subsection{Time Complexity}
Separating nodes based on repulsive forces has time complexity $ O(|N|^2) $ for each layer. 
For a complete network with $|N|$ nodes and $|L|$ layers, there are at most $( |L| \frac{|N| (|N|-1)}{2})$ in-layer edges and $ (\frac{|L| (|L|-1)}{2} |N|)$ inter-layer relationships in the whole network.
So, the time complexity of multiforce without using indexes is $O(|L| |N|^2 + |L|^2 |N|)$.

\begin{algorithm}
\caption{Multiforce}
\label{mylayout}
	\begin{algorithmic}[1]
\REQUIRE $G = (N, L, V, E)$: a multiplex network
\REQUIRE $W$: width of the frame
\REQUIRE $L$: length of the frame
\REQUIRE $\#\textrm{iterations}$
\STATE $f_r =$ function$(z,k) \{ $ return $k^2/z;$ \}
\STATE $f_a =$ function$(z,k) \{ $ return $z^2/k;$ \}
\STATE $\textrm{area} :=  W \cdot L$
\STATE $k :=  \sqrt{\frac{\textrm{area}}{|N|}};$
\STATE $t :=  \sqrt{|N|};$
\FOR { ($n \in N$) }
	\FOR { ($v \in V$ s.t.~$v$.node = $n$) }
		\STATE $(x,y) = $ random coordinates;
		\STATE $\textrm{pos}[v] = (x,y)$
		\STATE $z[v] := \textrm{index}(v.\textrm{layer});$
\ENDFOR
\ENDFOR
\FOR { ( $i=1$ to $\#\textrm{iterations}$ ) }
\STATE // calculate repulsive forces

\FOR{ ($v$ $\in$ $V$)} 
            \STATE disp[$v$] := $\vec{0}$;
            
            \FOR{ ($u$ $\in$ $V$)} 
                   
            \IF { ($u \neq v$ and $u$.layer = $v$.layer) }	
            

		\STATE $\bigtriangleup := \textrm{pos}[v] - \textrm{pos}[u];$

		\STATE disp[$v$] := disp[$v$] + $(\bigtriangleup /|\bigtriangleup|) \ast f_r(|\bigtriangleup|)$; 

		\ENDIF
\ENDFOR
\ENDFOR
           
         \STATE // calculate attractive forces inside each layer

\FOR { ($(u,v) \in E$) }


\STATE $ \bigtriangleup := \textrm{pos}[v] - \textrm{pos}[u];$

\STATE $\textrm{disp}[v] := \textrm{disp}[v] - (\bigtriangleup /|\bigtriangleup|) \ast f_a(|\bigtriangleup|,k) \ast \textrm{INLA}[v]$;
\STATE $\textrm{disp}[u] := \textrm{disp}[u] + (\bigtriangleup /|\bigtriangleup|) \ast f_a(|\bigtriangleup|,k) \ast \textrm{INLA}[u]$;

\ENDFOR  
 
   \STATE // calculate attractive forces across layers
 \FOR { ($n \in N$) }
	
 	\FOR { ($\{ u, v \}$ with $u, v \in V$, $u$.node = $v$.node = $n$) }
	
         \STATE $ \bigtriangleup := \textrm{pos}[v] - \textrm{pos}[u];$

         \STATE $\textrm{disp}[v] := \textrm{disp}[v] - (\bigtriangleup / |\bigtriangleup|) \ast f_a(|\bigtriangleup|,k) \ast \textrm{INTERLA}[v,u];$
         \STATE $\textrm{disp}[u] := \textrm{disp}[u] + (\bigtriangleup / |\bigtriangleup|) \ast f_a(|\bigtriangleup|,k) \ast \textrm{INTERLA}[u,v];$
	
	\ENDFOR
\ENDFOR

\STATE // assign new positions
\FOR{ ($v$ $\in$ $V$)} 
\STATE $\textrm{pos}[v] := \textrm{pos}[v] + ( \textrm{disp}[v] / | \textrm{disp}[v] |) \ast \min ( \textrm{disp}[v], t );$
\STATE $\textrm{pos}[v].x := \min(W/2, \max(-W/2, \textrm{pos}[v].x));$
\STATE $\textrm{pos}[v].y := \min(L/2, \max(-L/2, \textrm{pos}[v].y));$
\ENDFOR
\STATE // reduce the temperature
\STATE $t := \textrm{cool}(t);$
\ENDFOR
\end{algorithmic}
\end{algorithm}

\section{A simple example}
\label{sec:char}

Before providing a quantitative evaluation of our algorithm, we show the resulting layouts on a simple synthetic network and different weights, to give a visual intuition of it.
As we have briefly discussed in Section~\ref{sec:Related Work} there is a connection between aesthetic features of graph diagrams and their readability \cite{Moody2005}.
Typical metrics are the 
number of edge crossings, 
the number of overlapping nodes,
the separation of communities, and 
the representation of high-degree centrality nodes in a specific position, for example in the centre of each layer.
These features can be easily manually inspected in the following diagrams. In addition, the simple network used in this section allows us to check the impact of different settings for the in-layer and inter-layer weights.

 \begin{figure*}[htp]
\caption{The structure of the synthetic network}
\label{datasetstructure}
\begin{center}
 \begin{tabular}{ | m{5cm}| m{5cm}|}
 \hline 
{First Layer}&{Second Layer}
 \\
 \hline 
 
 \raisebox{-\totalheight}{\includegraphics[width=0.20\textwidth, height=30mm]{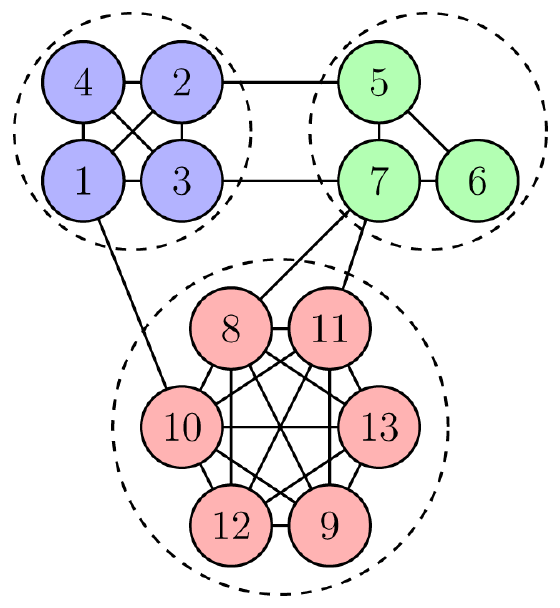}}
  & \raisebox{-\totalheight}{\includegraphics[width=0.25\textwidth, height=30mm]{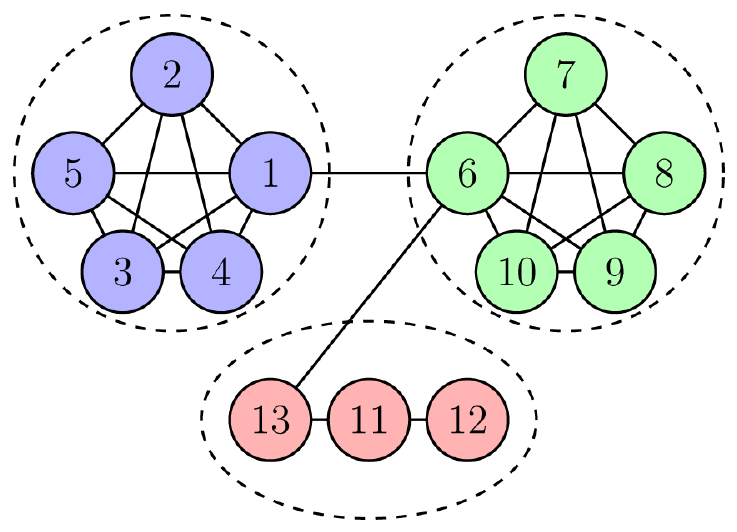}}
\\ 
\hline 

 \end{tabular} 
 \end{center}
 \end{figure*}

The structures of the synthetic dataset used in the following is shown in Figure \ref{datasetstructure}.
This dataset 
contains two layers, each with 13 nodes; the whole network has 53 in-layer edges. Some nodes that are present in one community in one layer belong to another in the second layer, making this small example useful to show how these nodes are handled by varying the weights.

 \begin{table*}[htp]
\caption{Visualizing the synthetic network using different weights}
\label{differentlayertable}
\begin{center}
\begin{tabular}{ | m{4em} | m{4cm}| m{4cm} | m{4cm}|}
 \hline 
{} & {Inter-layer Force=0} &{Inter-layer Force=0.5} 
 & {Inter-layer Force=1}  \\
 \hline 
 {In-layer Force=0} &  \raisebox{-\totalheight}{\includegraphics[width=0.2\textwidth, height=30mm]{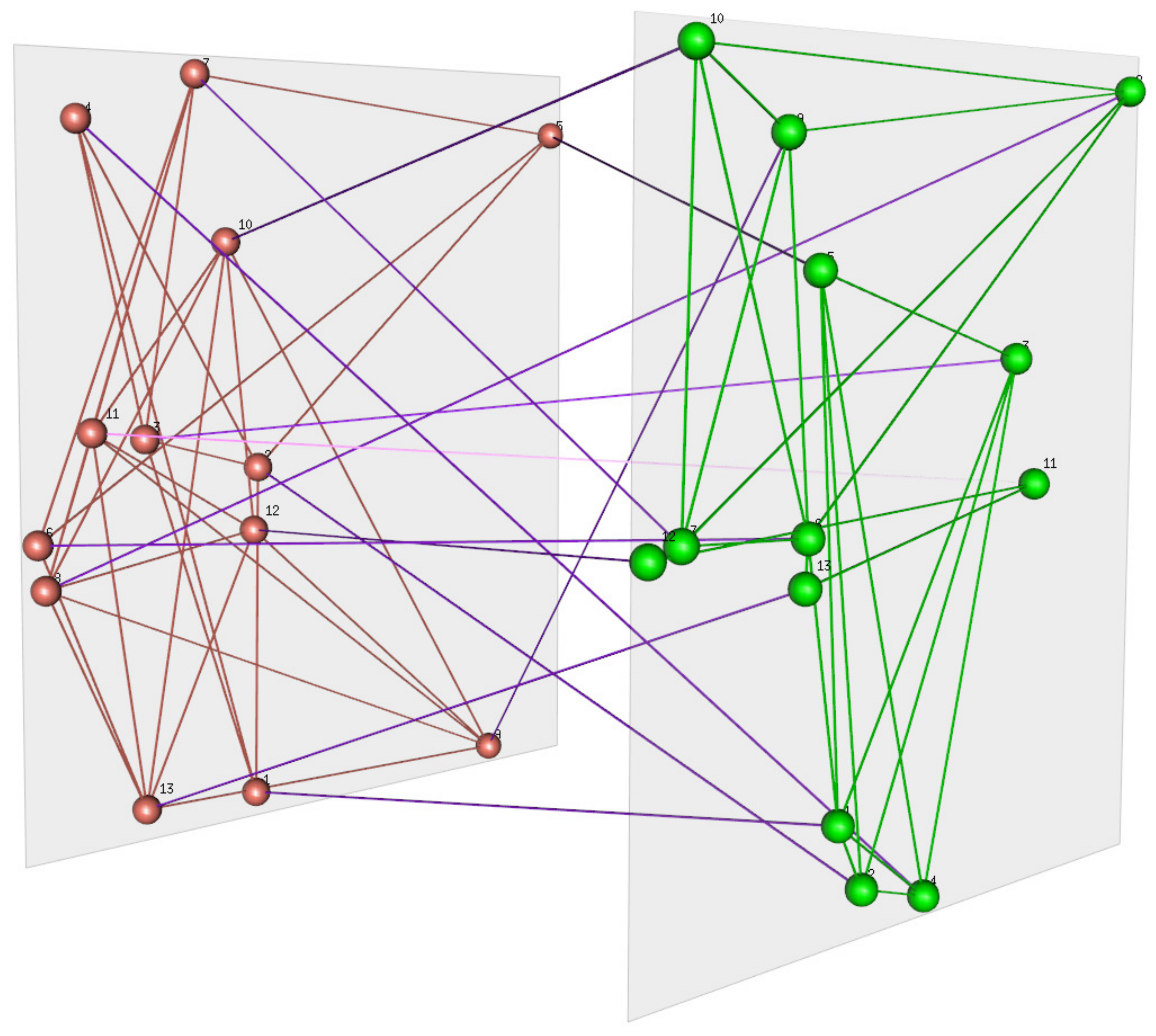}}  { 
 

} & \raisebox{-\totalheight}{\includegraphics[width=0.2\textwidth, height=30mm]{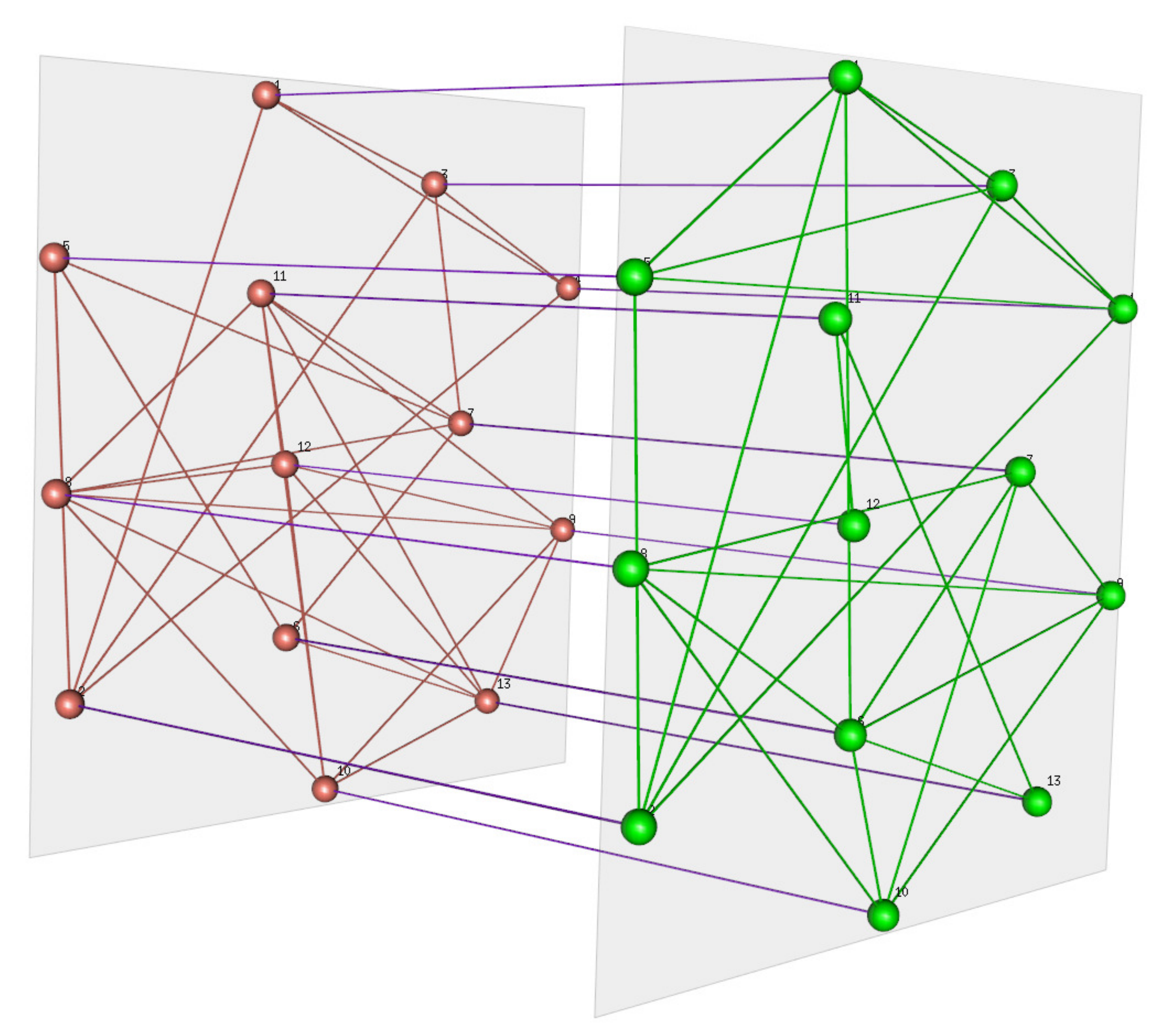}}  {

 
} & \raisebox{-\totalheight}{\includegraphics[width=0.2\textwidth, height=30mm]{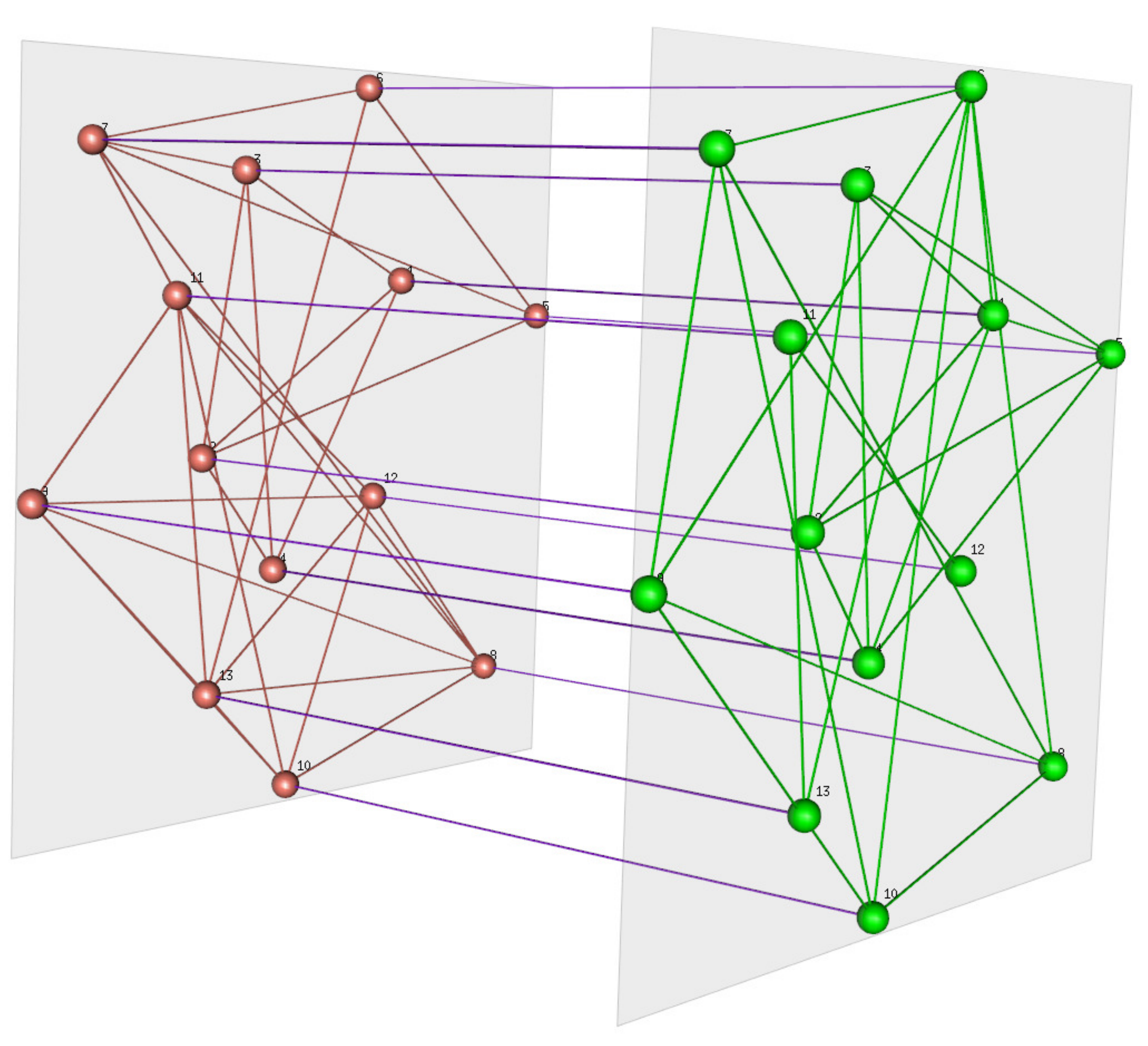}}{ 

 
}
\\ 
\hline 
{In-layer Force=0.5} &\raisebox{-\totalheight}{\includegraphics[width=0.2\textwidth, height=30mm]{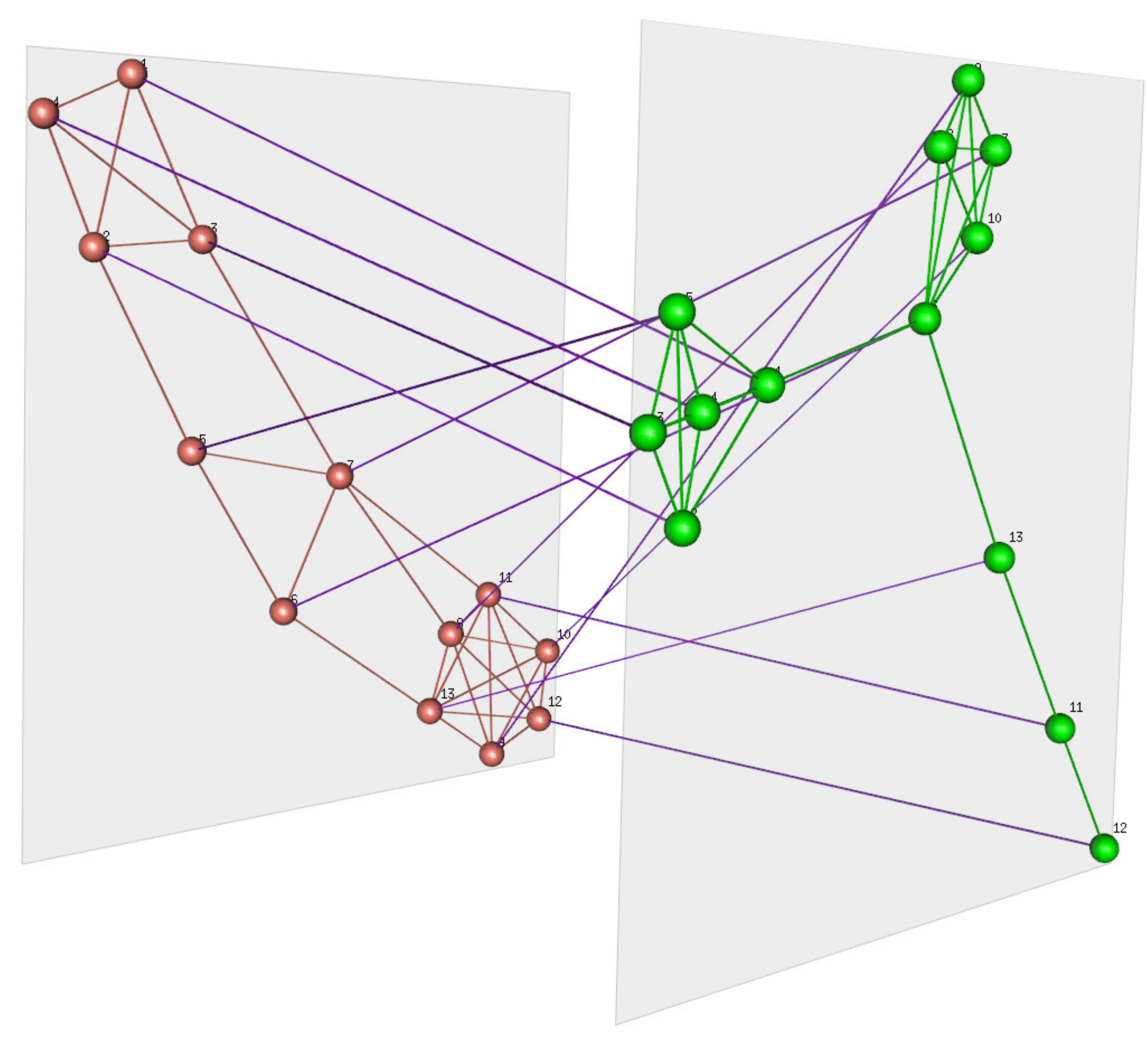}} {


} & \raisebox{-\totalheight}{\includegraphics[width=0.2\textwidth, height=30mm]{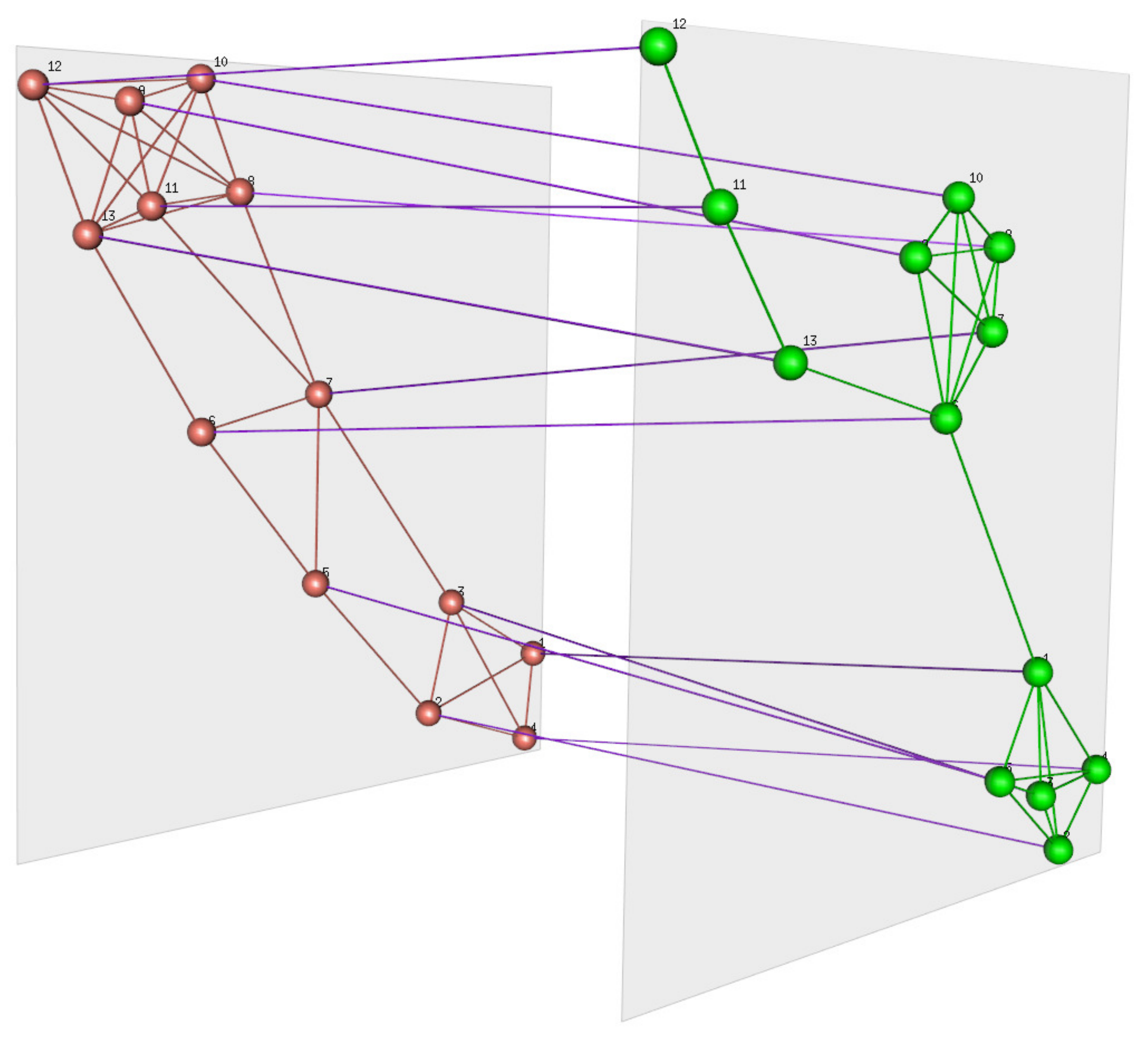}}{

 
} &\raisebox{-\totalheight}{\includegraphics[width=0.2\textwidth, height=30mm]{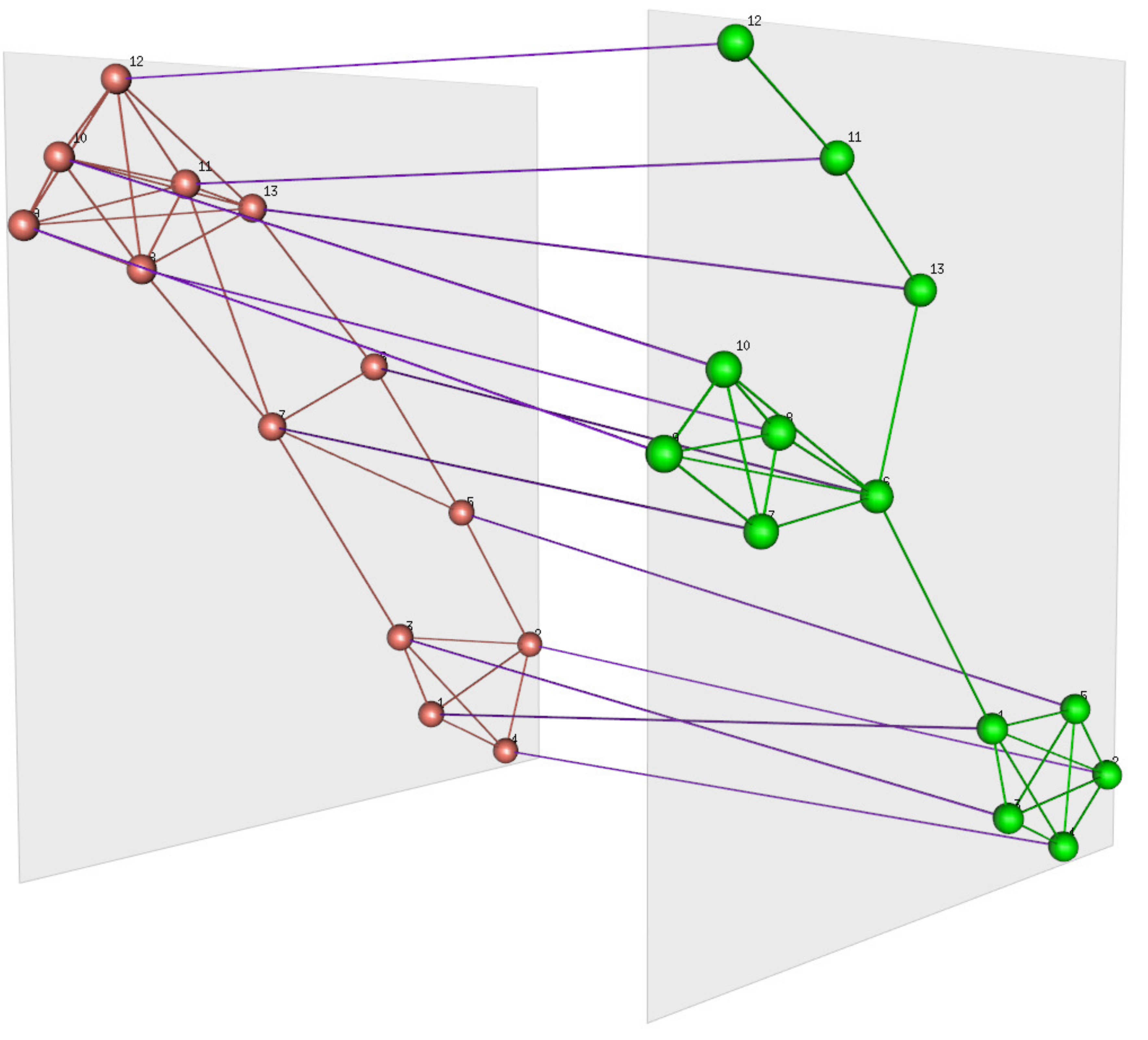}}{ 

 
}
\\ 
\hline 
{In-layer Force=1} &  \raisebox{-\totalheight}{\includegraphics[width=0.2\textwidth, height=30mm]{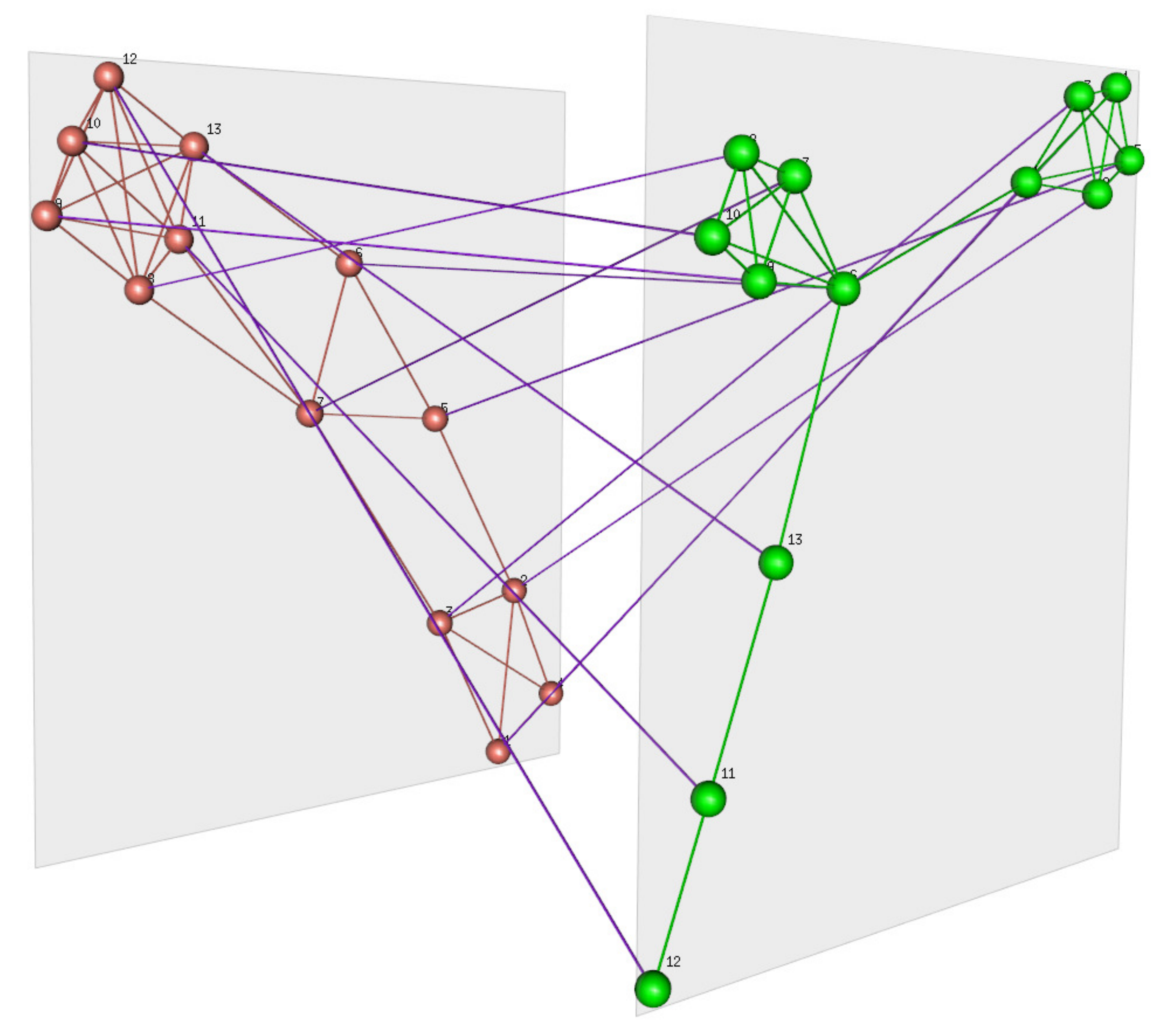}} {


} & \raisebox{-\totalheight}{\includegraphics[width=0.2\textwidth, height=30mm]{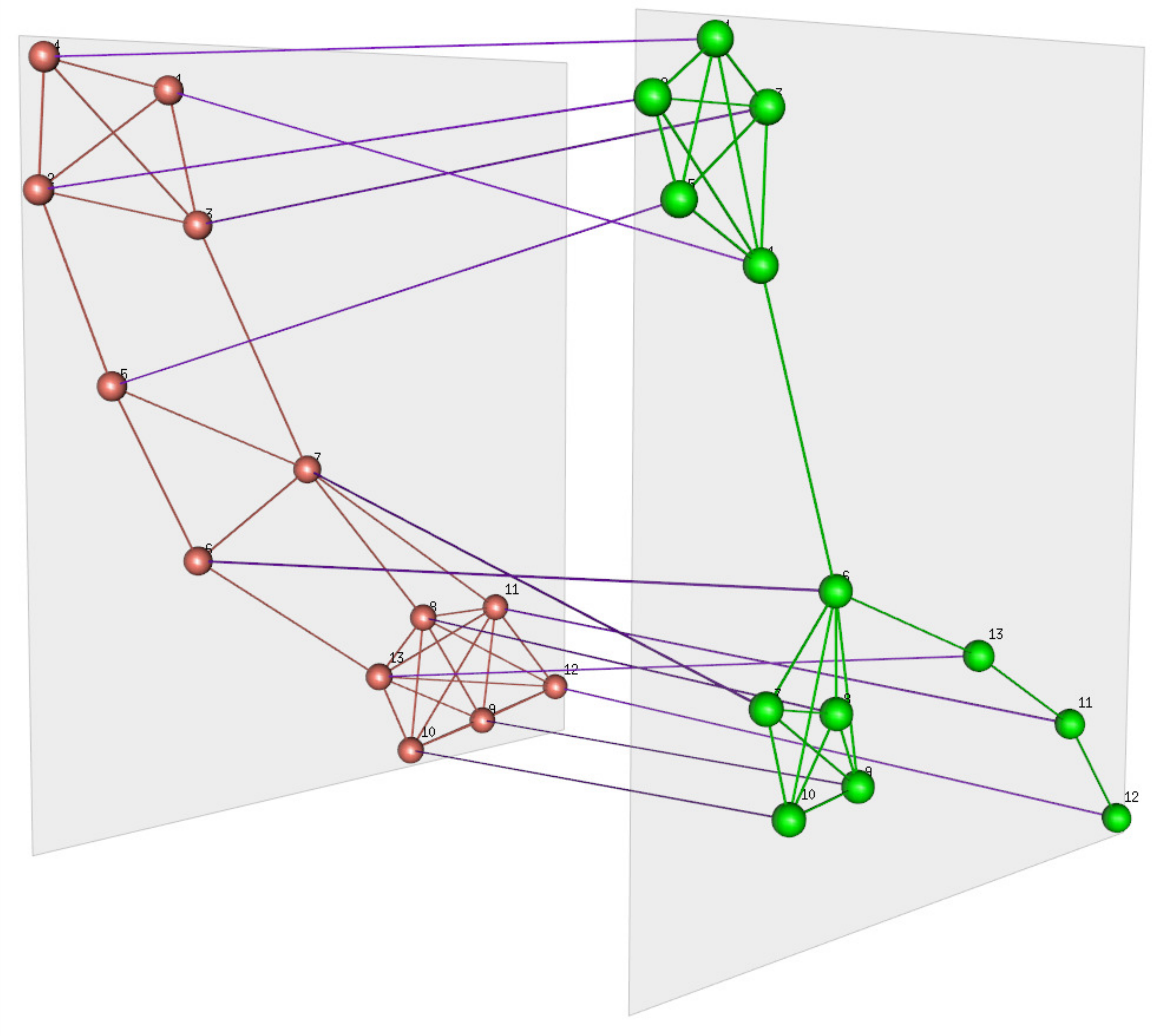}}  {

 
} & \raisebox{-\totalheight}{\includegraphics[width=0.2\textwidth, height=30mm]{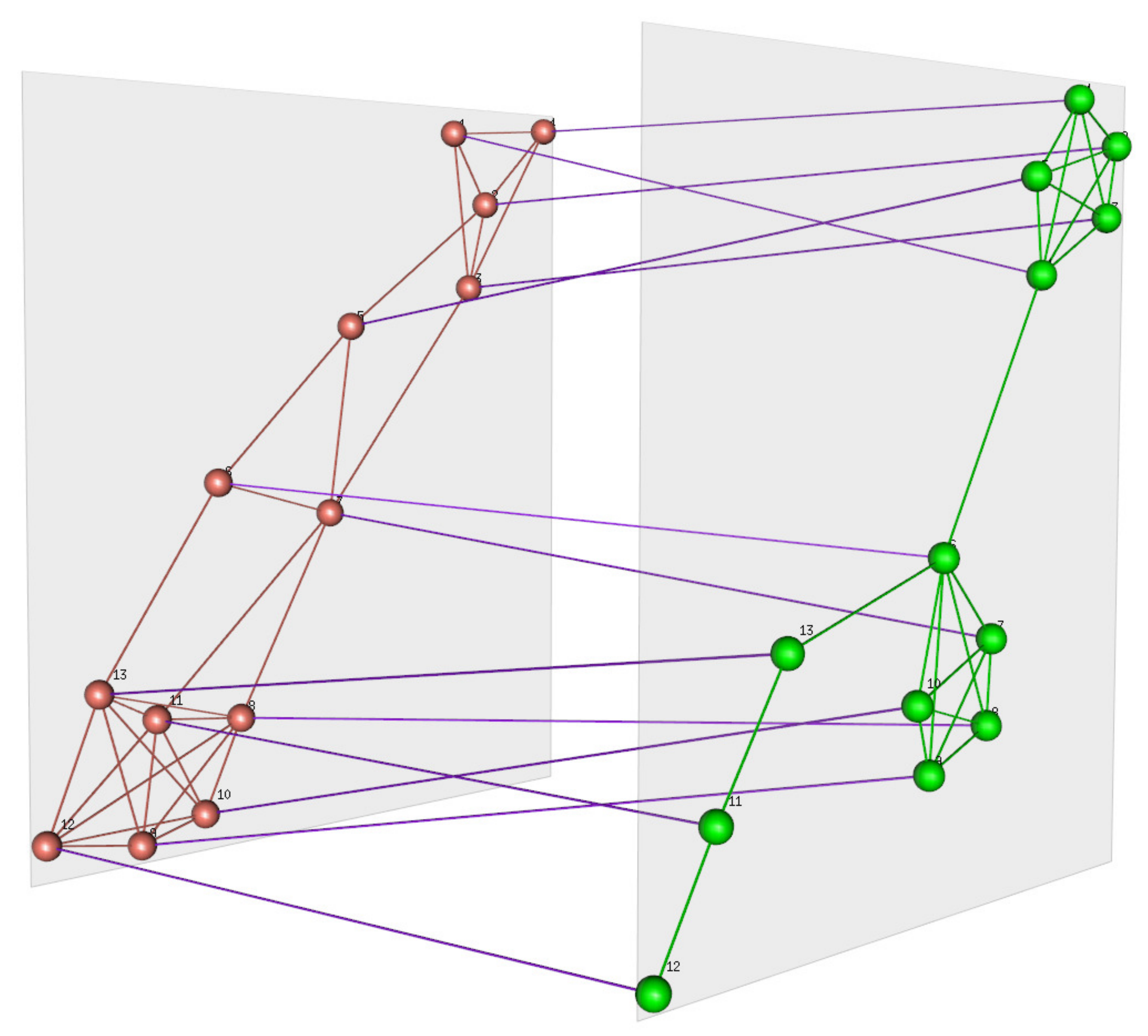}} { 

 
}
\\ 
\hline 

  \end{tabular} 
 \end{center}
 \end{table*}

The results of drawing this multiplex network with multiforce and different combinations of weights are shown in Table  \ref{differentlayertable}.
While this is just a first illustrative example it highlights the main features of the proposed method, later tested with a larger qualitative example on real data and a quantitative evaluation.
\begin{itemize}
\item When a good compromise between in-layer and inter-layer forces can be found, as in this simple example, the method is stable with respect to its parameters. All the examples in Table  \ref{differentlayertable} where both forces are not null (that is, the four bottom-right plots) are very similar.
\item If we focus on the bottom-right plot, we can see how despite some differences in structure most of the nodes are aligned, with the nodes being part of different communities in the different layers having slightly different positions that preserve the local structures. In other words, we can highlight the local structures in both layers at the same time, still keeping the ability to easily identify corresponding nodes across layers.
\item If we extend our analysis from single nodes to whole structures, such as communities or sub-graphs, an inspection of the inter-layer connections may reveal which parts of the layers are similar and which parts are significantly different. We will in fact find similar structures well aligned across different layers, with approximately parallel inter-layer connections among nodes. Dissimilar parts of the network will instead be characterized by more oblique and cahotic inter-layer connections (if in-layer forces are prevalent, preventing alignment across layers) or meaningless/random layouts (if inter-layer forces are prevalent, removing local structures) or combinations of these two cases. Notice that this is even better than what we can do using existing layer correlation metrics (such as degree-degree correlation or the jaccard coefficient), which compare whole layers and are thus not able to identify locally aligned regions.
\end{itemize}

\section{Experimental evaluation}
\label{sec:eval}

In this section we present a more structured evaluation of our method, using real datasets.

Our algorithm does not replace existing slicing approaches: it unifies and complements them with new options, given by the recognition of the existence of in-layer and inter-layer forces and the introduction of balancing parameters between them. In theory, all the available options can be valuable, including those provided by existing methods: for a user it can be good to keep all the nodes strictly aligned, while for another it can be better to visualize the layers one by one.
Therefore, we will not try to prove the superiority of one approach over the other, and the main value of our proposal is to have a single flexible algorithm that can emphasize different aspects of the data. However, we can also unambiguously show that our algorithm can generate valuable visualizations that cannot be obtained using existing methods. By being less strict and, for example, allowing some small disalignment between nodes in different layers, still not preventing their quick identification, we can get the best from both extremes.

More precisely, we compare different layouts against an ideal case where each layer is drawn independently of the others, to optimize its internal readability, and at the same time all occurrences of the same node across different layers are perfectly aligned. Notice that such a visualization is impossible to obtain in general, except when the two layers are (almost) identical.
To compute the distance between the ideal case and our tests we can define two measures representing these two criteria: the sum of the
forces still active on the nodes inside each layer (called internal fit) and the displacement between nodes in different layers (caled external fit), also computed as the sum of the inter-layer forces still active at the end of the algorithm.
Notice that an optimal diagram as defined above would contain both internally stable and aligned nodes, so both measures would be 0, indicating the best possible layout (assuming that a force-based approach is used).
The hypothesis we test in the following experiments is that existing approaches optimize only one of these two criteria, while multiforce can obtain good scores on both at the same time when using both in-layer and inter-layer forces. In the following we call this setting, allowed by our method, \emph{balanced}.

\begin{figure}
\centering
\includegraphics[width=.5\textwidth]{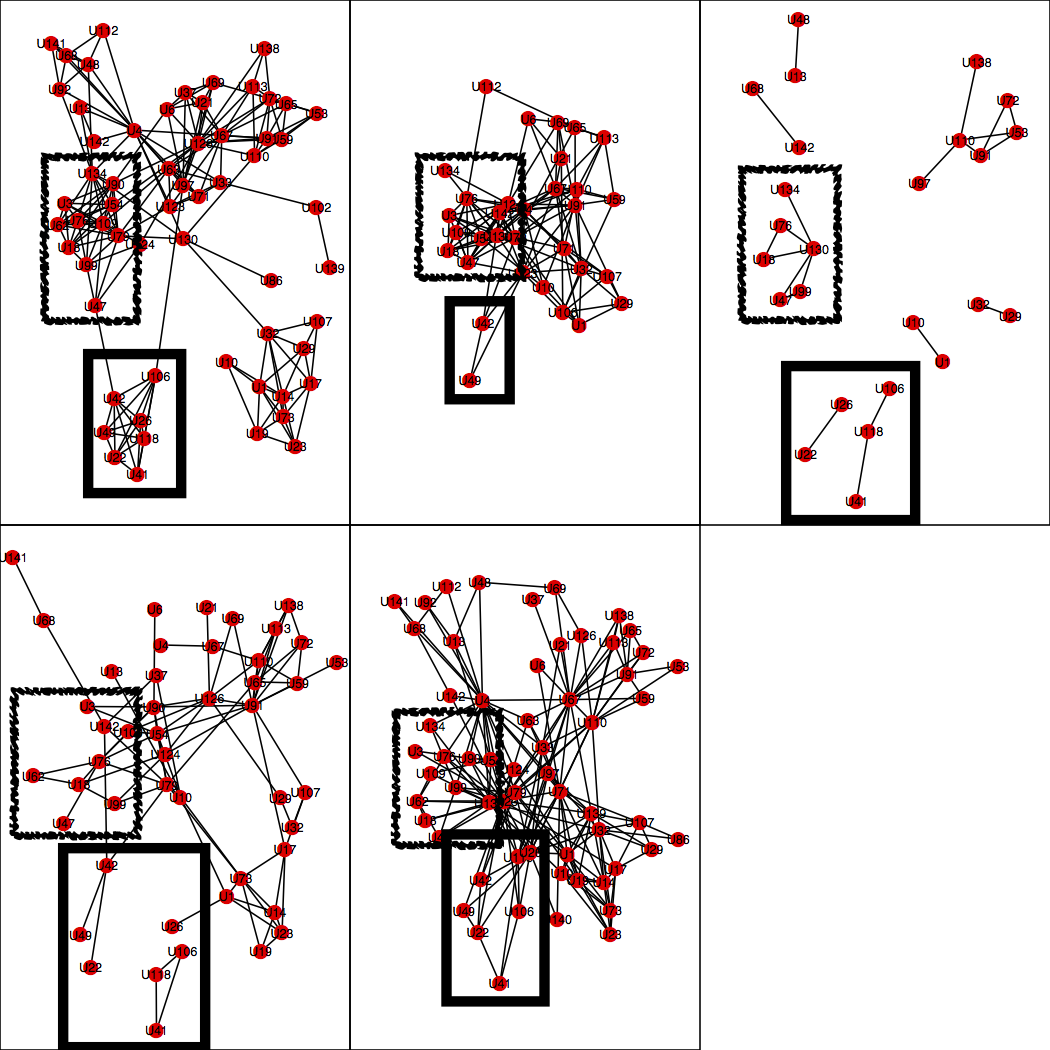}
\caption{Alignment of network structures across layers using a balanced visualization, with both in-layer and inter-layer forces active}
\label{fig:ml}
\end{figure}

To give an intuition of how a balanced drawing may look like, consider Figure \ref{fig:ml} where we have visualized a real 5-layer social network previously used in multiplex visualization research \cite{2015arXiv150101666M}. To make the diagram more readable on paper we have drawn each layer besides the others instead of in a 2.5D space. The different layers are drawn according to their internal organization, showing peculiar structures: for example, the lower community indicated in the top-left diagram (continuous line) is not present in some of the other layers. However, we can see how the nodes belonging to this community have been visualized at similar locations in the other layers, providing information about inter-layer relationships. We can see, for example, that some of the nodes in this community are not present in the second layer, that this community is split into two sub-communities in the third layer, that it is not present in the fourth layer, where the same nodes are connected to different parts of the graph, and that they form a similar but less dense community in the fifth layer. Notice that in the fourth layer the nodes are still more or less in the same position as in the first layer, despite the fact that they do not form a community. In this way it becomes easier to locate them -- in the real 2.5-dimensional visualization we would also have lines connecting nodes across layers, as in the previous section, making the task straightforward as long as the nodes are more or less aligned. In the figure we have also marked a second community with a fuzzy rectangle, so that the reader may observe another example.

To quantitatively support our claim we have executed the algorithm on eight real datasets from different domains, summarized in Table~\ref{datasetstructure1}: 
\begin{itemize}
\item
\textbf Three traditional multiplex networks from the the Social Network Analysis literature.
\item A hybrid online/offline social network
with five types of relationships among the employees of a Computer Science Department
\cite{2015arXiv150101666M}.
\item A dataset showing the relationship among physicians in four different cities, collected to investigate information diffusion about drugs
\cite{10.2307/2785979}. 
\item A criminal network \cite{Roberts2011}.
\item A biological network in which every layer shows a synaptic junction (Electric, Chemical Monadic, Chemical Polyadic) \cite{Chen21032006}. 
\item A transport network with flight connections in Europe \cite{Cardillo2013}.
\end{itemize}

  \begin{table*}[htp]
\caption{Properties of Real-world Networks}
\label{datasetstructure1}
\begin{center}
 \begin{tabular}{ | l | l | l | l |}
 \hline 
{Network} &{\# Layers}&{\# Nodes}&{\# Edges}
 \\
\hline 
\hline 
{Social -- Padgett} & {2} & {15}&{35}
\\ 
\hline 
{Social -- Kaptail} & {4} & {39}&{552}
\\ 
\hline 
{Social -- Wiring} & {6} & {14}&{108}
\\ 
 \hline 
{Hybrid -- AUCS} & {5}& {61}&{1240}
\\ 
\hline 
{Social -- Physicians} & {3} & {241}&{1370}
\\ 
\hline 
{Criminal -- Noordin} & {4} & {74}&{318}
\\ 
\hline 
{Biological -- synapses} & {3} & {279}&{3108}
\\ 
\hline 
{Transport -- airlines} & {37} & {417}&{3588}
\\ 
\hline 
  \end{tabular} 
 \end{center}
 \end{table*} 

\begin{figure*}
\centering
\subfigure[Social -- Padgett]{\label{fig:2-4}\includegraphics[width=.3\textwidth]{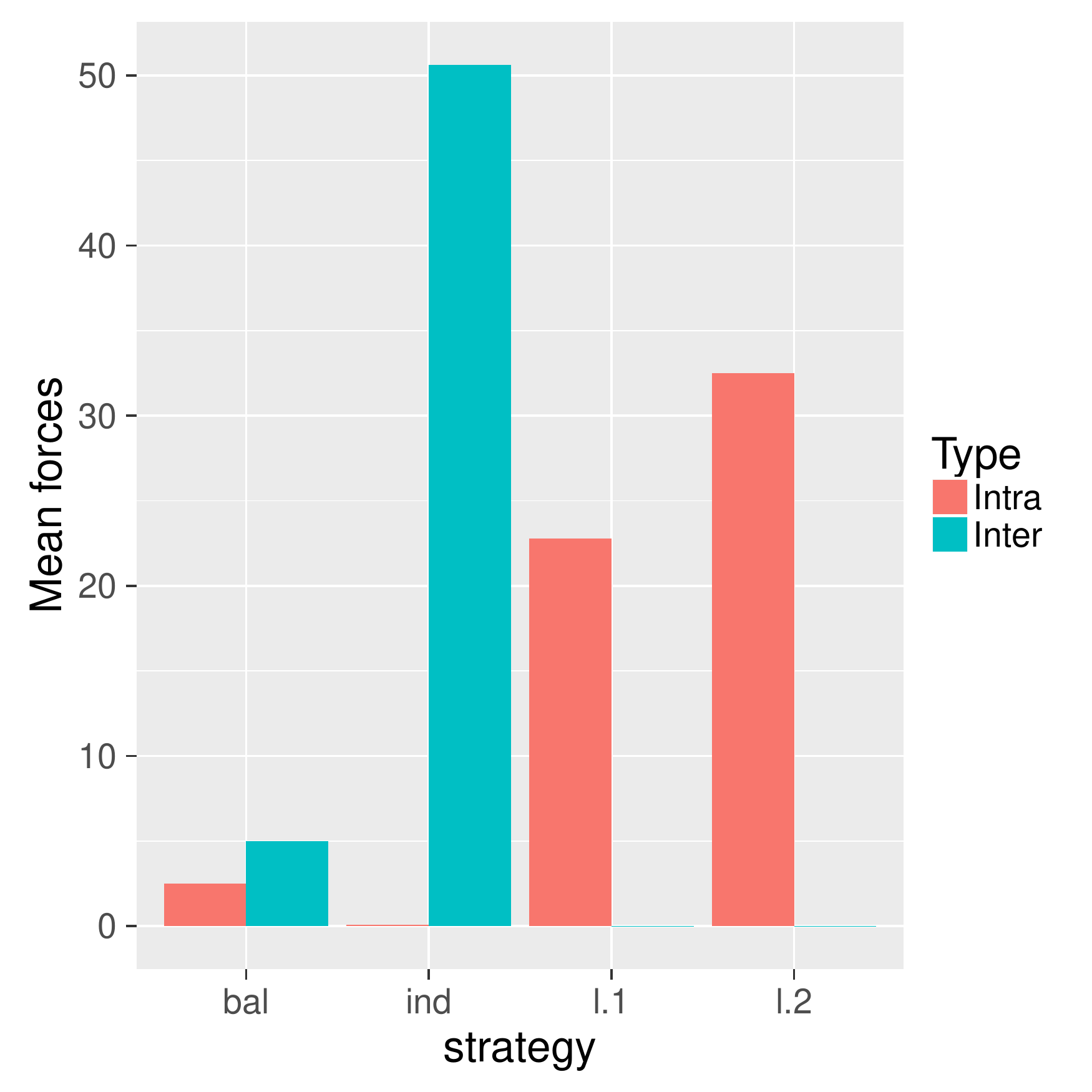}}
\subfigure[Social -- Kaptail]{\label{fig:2-5}\includegraphics[width=.3\textwidth]{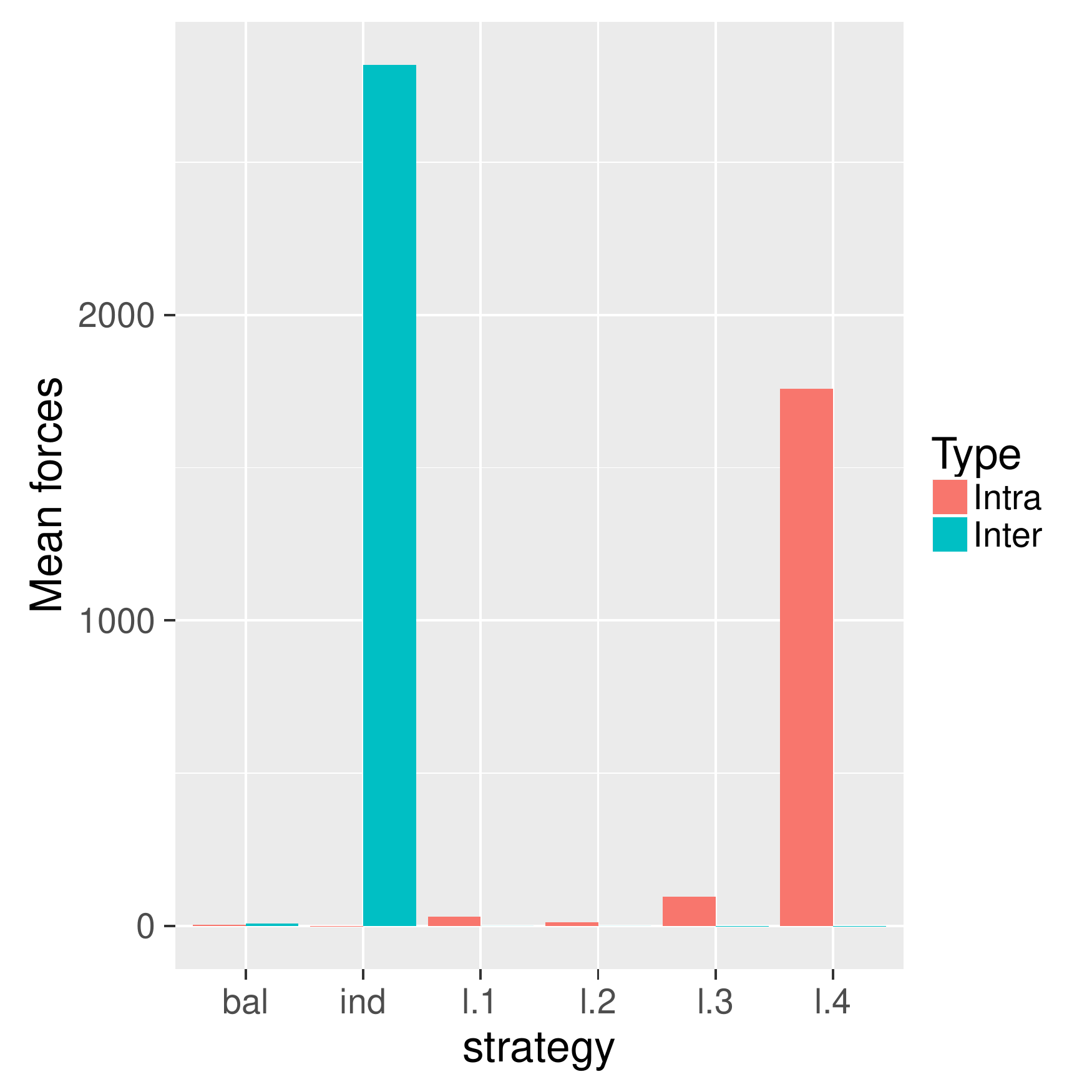}}
\subfigure[Social -- Wiring]{\label{fig:2-6}\includegraphics[width=.3\textwidth]{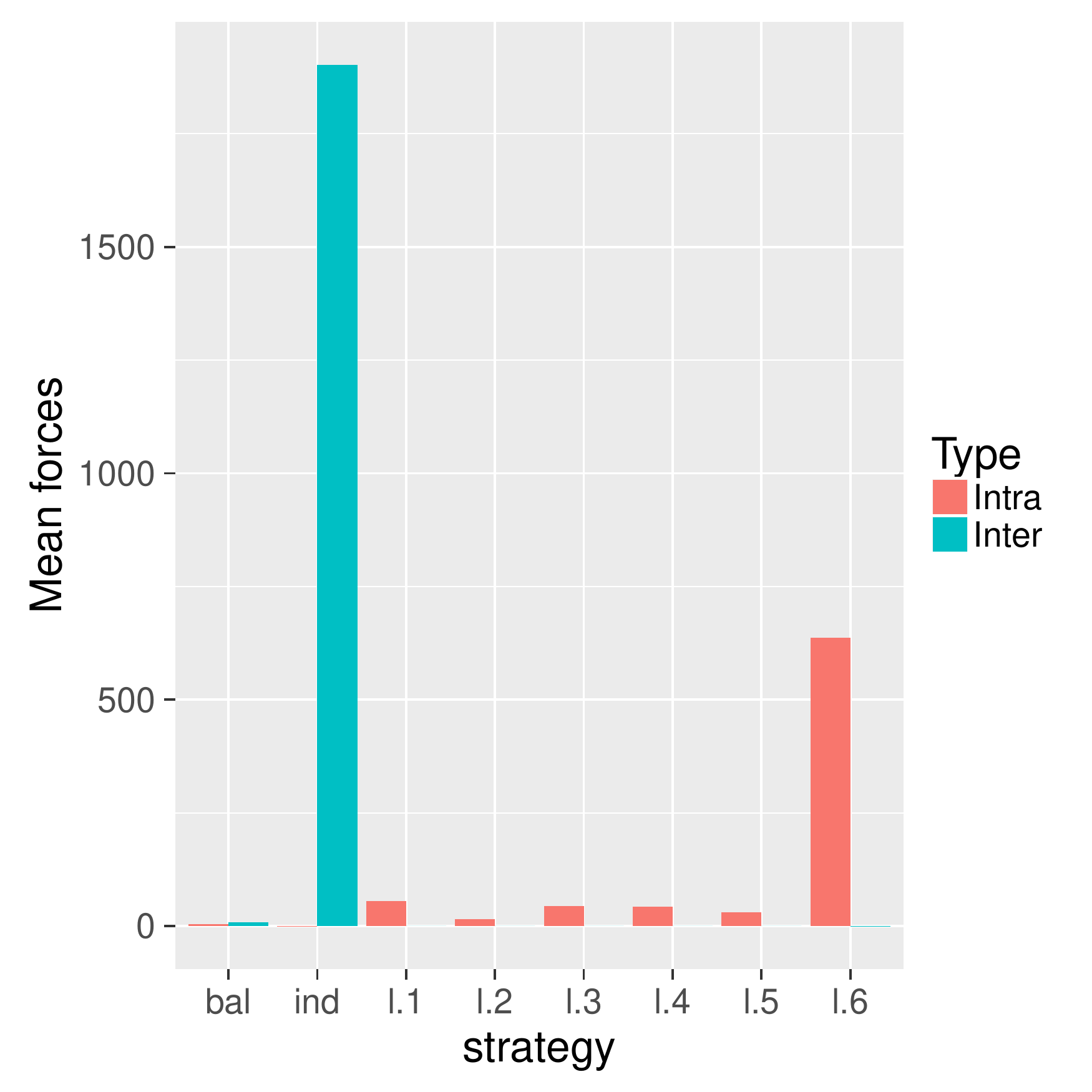}}
\subfigure[Hybrid -- AUCS]{\label{fig:2-1}\includegraphics[width=.3\textwidth]{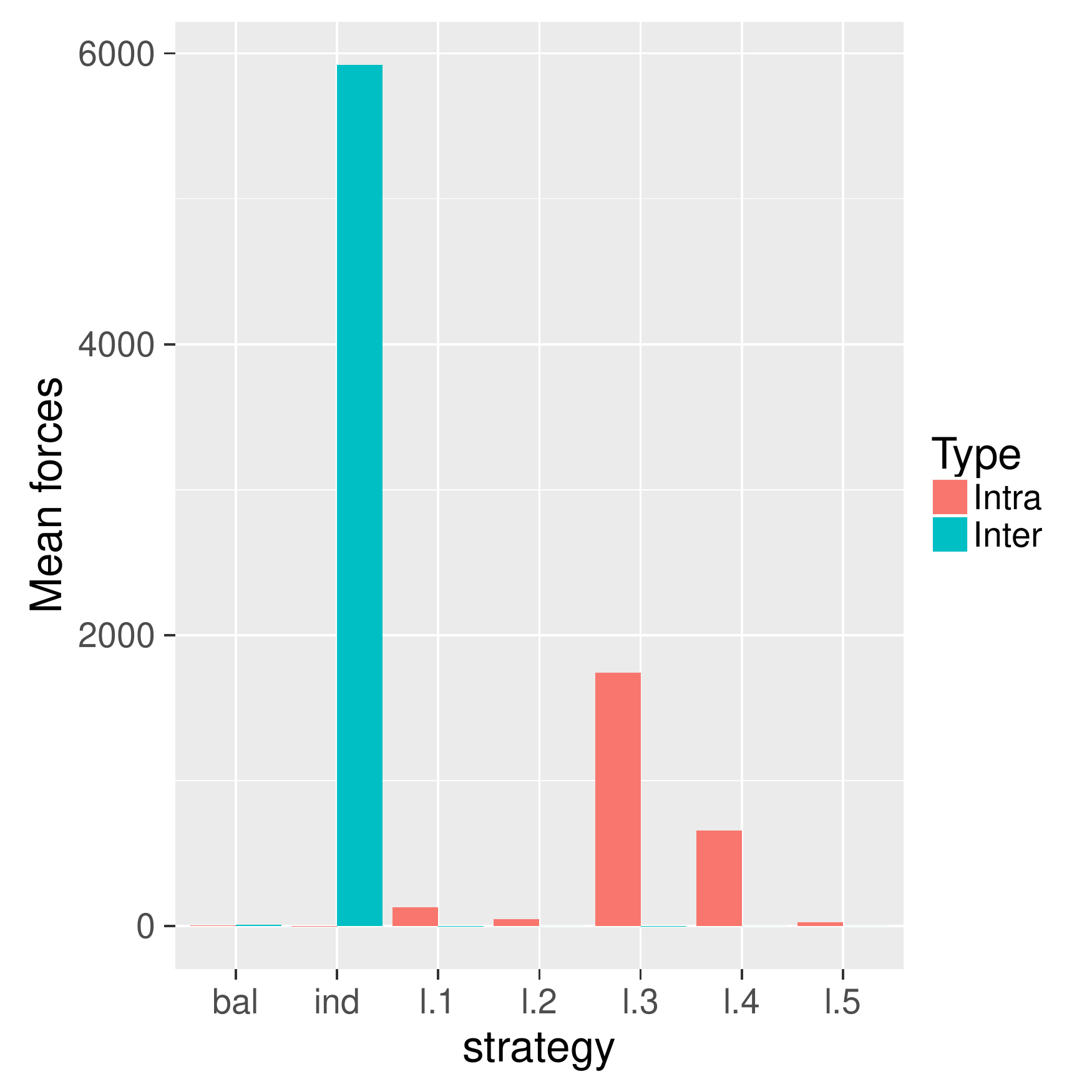}}
\subfigure[Social -- Physicians]{\label{fig:2-7}\includegraphics[width=.3\textwidth]{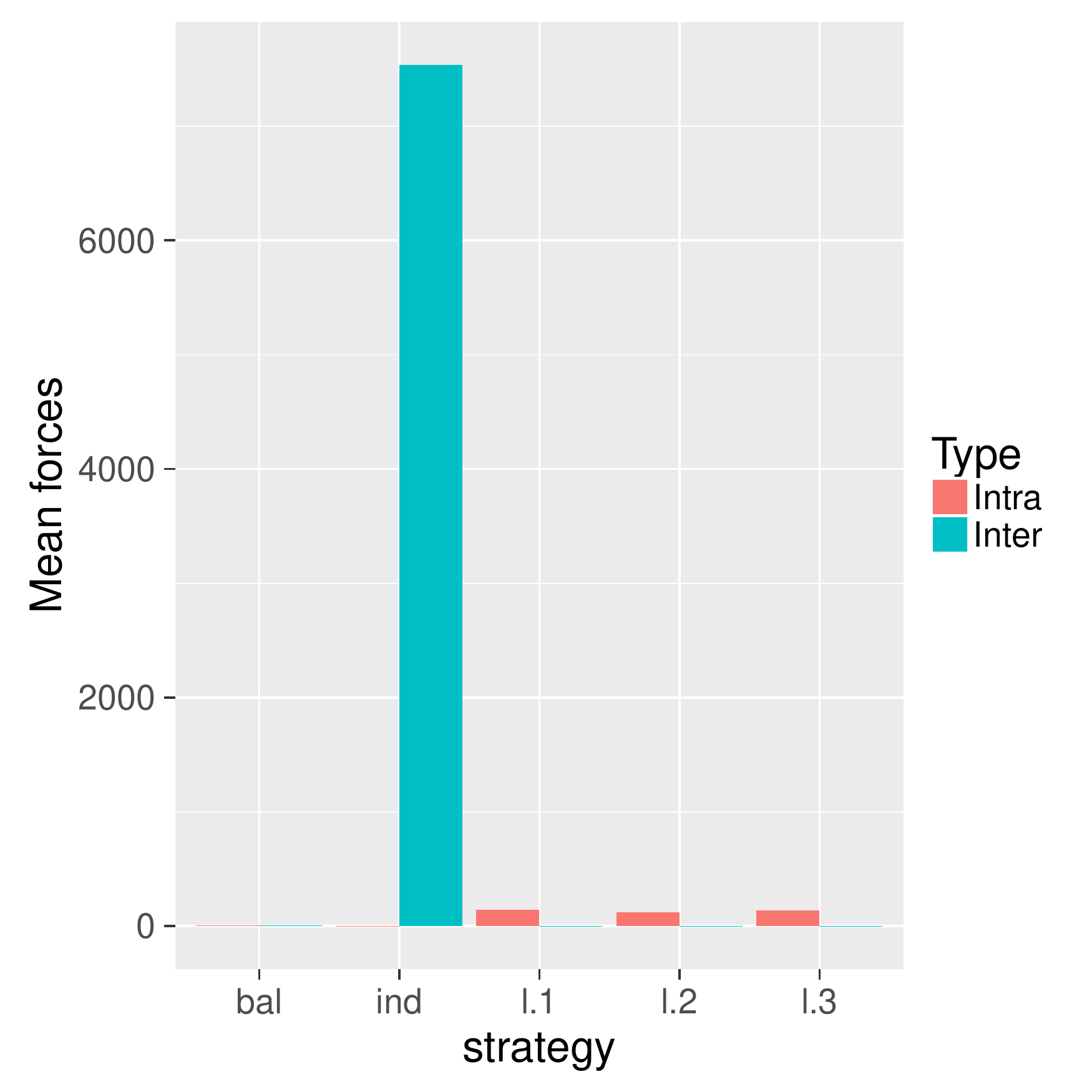}}
\subfigure[Criminal -- Noordin]{\label{fig:2-3}\includegraphics[width=.3\textwidth]{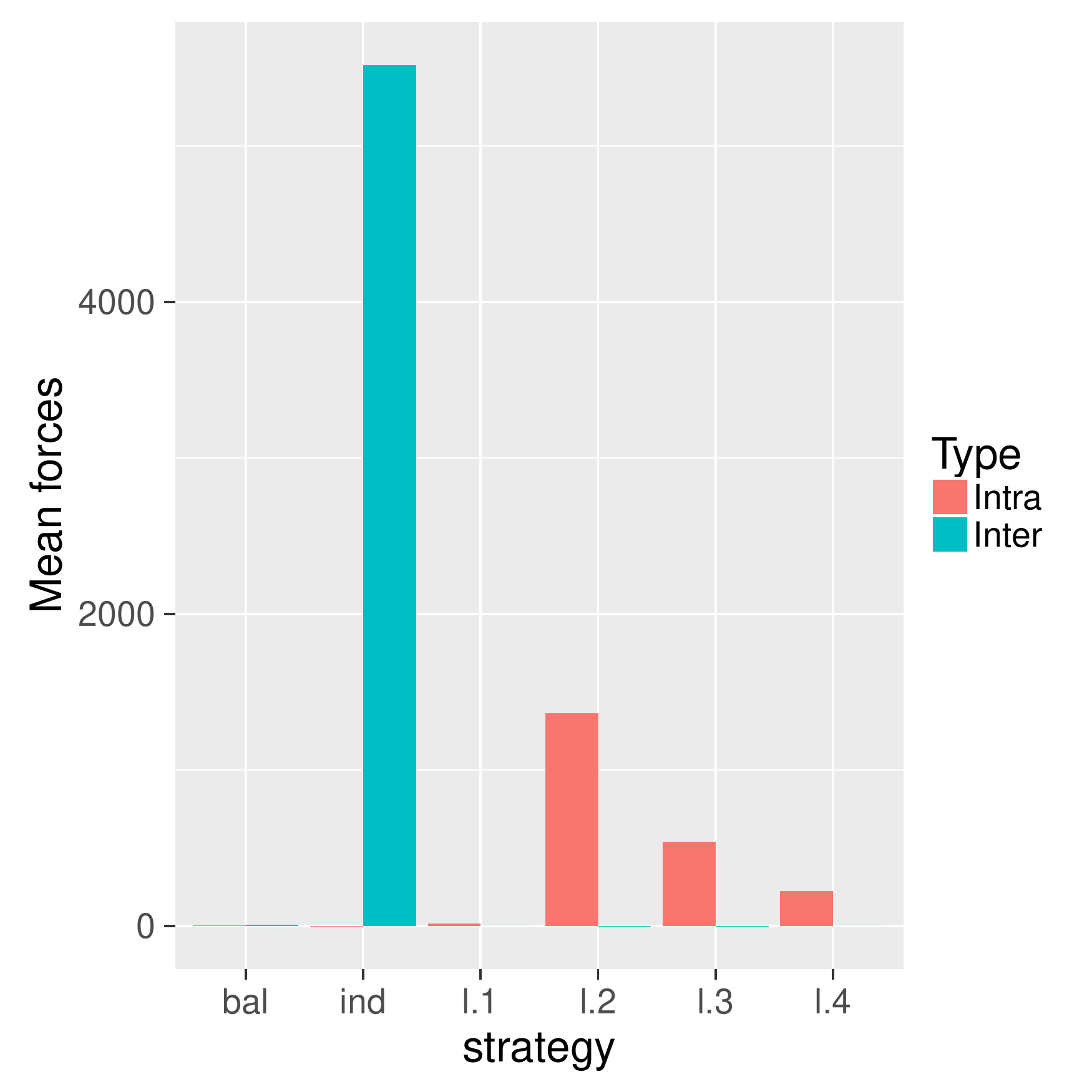}}
\subfigure[Biological -- synapses]{\label{fig:2-8}\includegraphics[width=.3\textwidth]{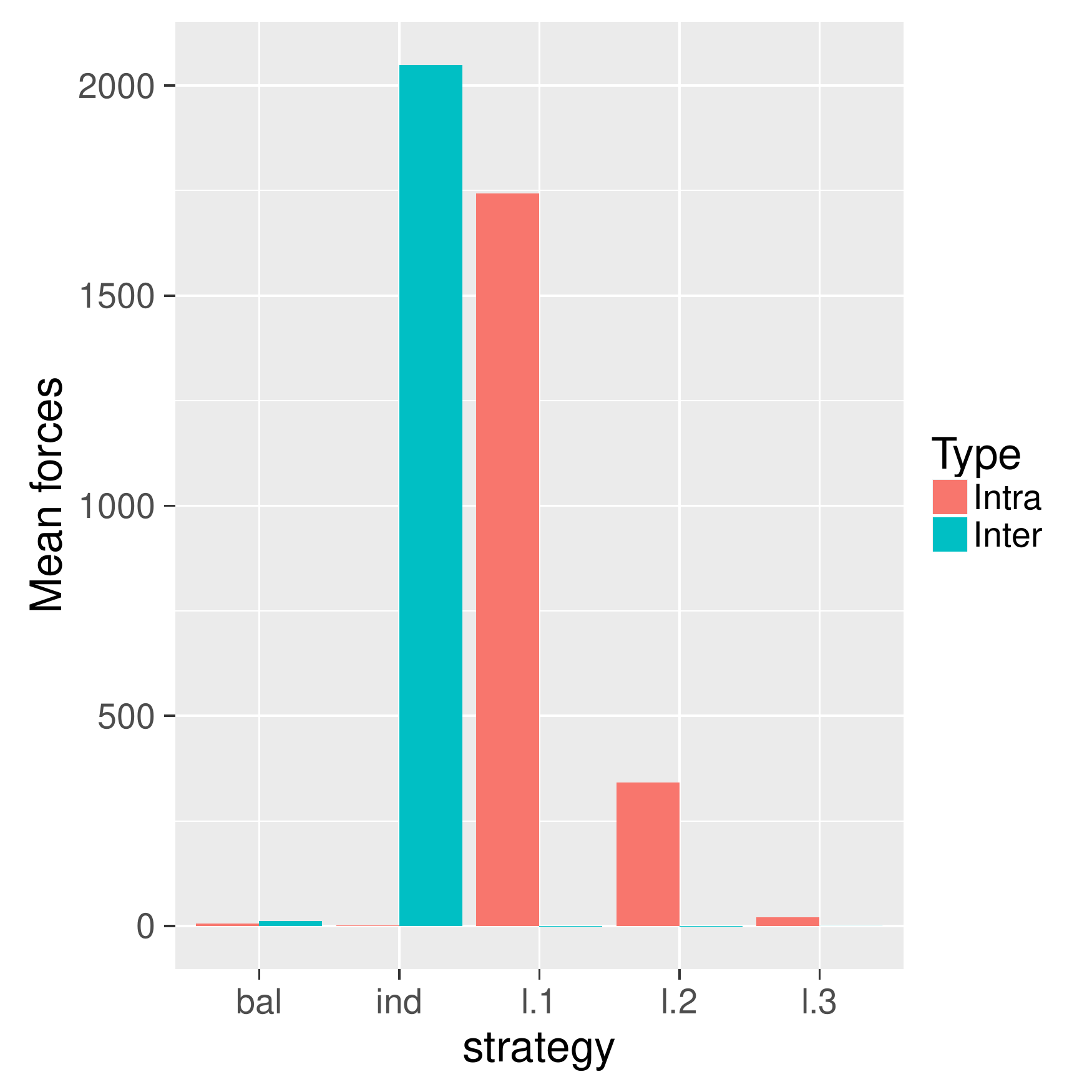}}
\subfigure[Transport -- airlines]{\label{fig:2-2}\includegraphics[width=.3\textwidth]{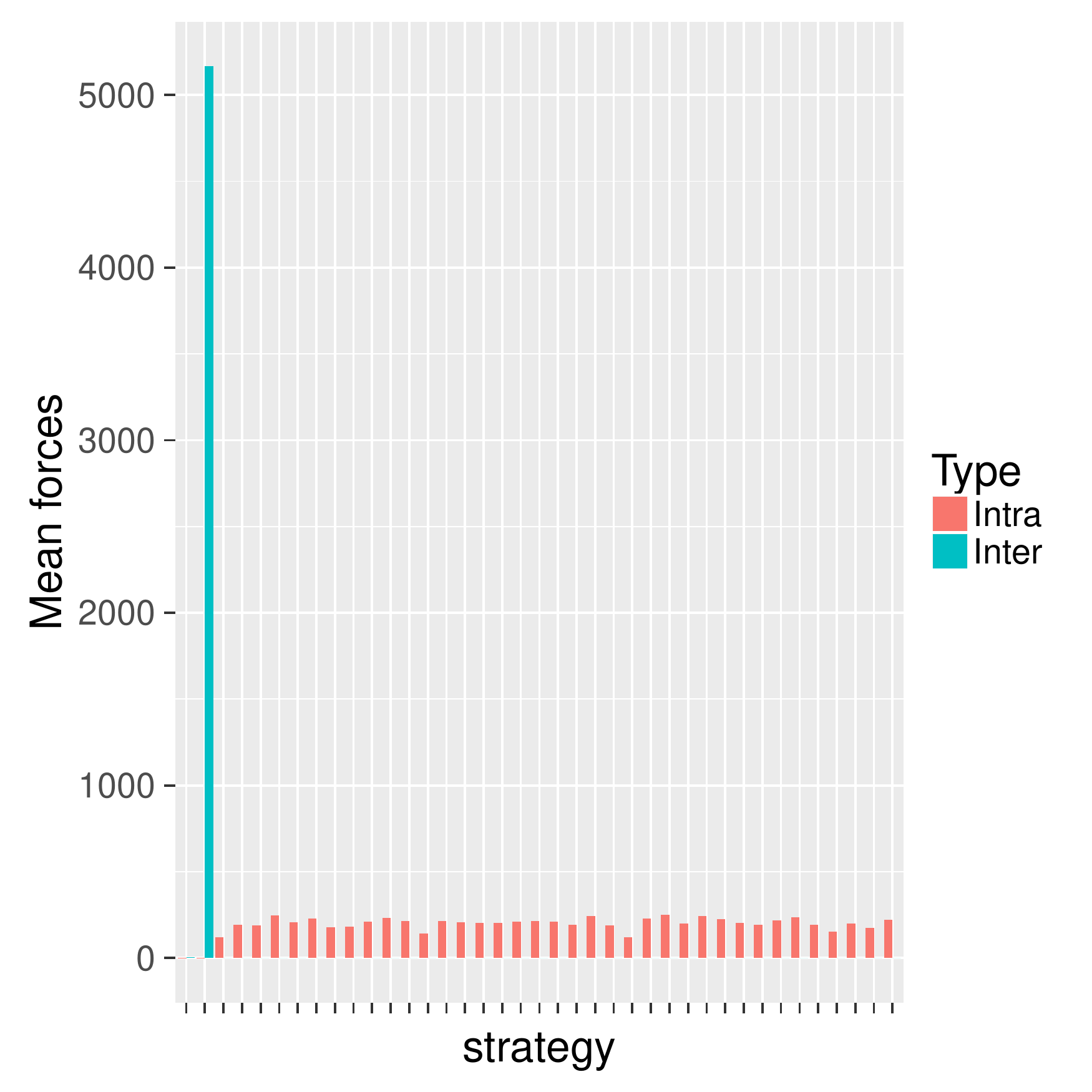}}
\caption{Results of experiments on real data}
\label{fig:two criteria}
\end{figure*}

For each network, Figure \ref{fig:two criteria} shows how different layouts behave with respect to the two aforementioned measures, indicated as Intra (internal fit) and Inter (external fit) in the plots. We remind the reader that high values of the former means that some layers have not been correctly visualized according to their internal structure --- for example, the nodes in a community may have been spread around the frame instead of being close to each other. A high value of the latter means that nodes are not aligned across layers --- for example, a node may have been visualized in the top left corner in one layer and in the top right in the other. The ideal case would be to have 0 for both metrics. In each figure, several settings are tested, each represented by two bars corresponding to the two metrics. The first setting corresponds to a balanced layout, where our algorithm is executed with weight 1 for all inter- and in-layer forces. The second setting corresponds to the case where all layers are visualized independently of the others. In the remaining settings the layout is computed based on one of the layers, and nodes are kept aligned on the other layers.

As expected, for each network (that is, for each plot in Figure \ref{fig:two criteria}) the independent visualization (second case) generates nodes that are not well aligned, represented by a tall second bar, and in the following cases we can see that computing the layout based on one layer prevents other layers from having good internal layouts, as shown by the tall first bars. The balanced option (first case) presents internal layouts that are less good than the ideal case, and node alignments that are also less good than the best possible option, but both are close to optimal and significantly better than the aspect not optimized in other experiments. In practice, this corresponds to layouts similar to the one shown in Figure \ref{fig:ml}.

\section{Conclusion}
\label{sec:Conclusion}
An optimal layout for multiplex networks would be able to reveal the structure of each layer and the relationships between different layers at the same time. Unfortunately, this is not possible in general, because these two aspects may not be aligned in real data, with some layers being very uncorrelated with the others.

To address this problem, we proposed multiforce, a force-directed algorithm in which both in-layer and inter-layer forces can affect the position of nodes.
In-layer forces keep together connected nodes and improve the identification of communities, while inter-layer forces help users finding the same nodes in different layers. 

In the evaluation of the method we showed that while the algorithm supports more traditional layouts it can also generate what we call \emph{balanced visualizations} where both internal properties and node alignments are handled. This has been presented on a simple synthetic example, to highlight the main features of our method, on a real dataset, to give a qualitative idea of how the layouts produced by this approach look like, and quantitatively, introducing two quality metrics and showing how a balanced approach can satisfactorily address both at the same time on several real datasets from different domains.

In this paper we only considered static multiplex networks, but nodes and edges can change over time and appropriate layouts considering dynamical features could be valuable tools. In addition, other aesthetic features of graph diagrams like symmetry and uniform edge length could be investigated in the context of multiplex networks.
Finally, other types of inter-layer links can also be considered in the future, as in more general multilayer network models.


\begin{thebibliography}{10}

\bibitem{gravity}
Michael~J. Bannister, David Eppstein, Michael~T. Goodrich, and Lowell Trott.
\newblock Force-directed graph drawing using social gravity and scaling.
\newblock In {\em Proceedings of the 20th International Conference on Graph
  Drawing}, pages 414--425. Springer-Verlag, Berlin, Heidelberg, 2013.

\bibitem{battiston2014structural}
Federico Battiston, Vincenzo Nicosia, and Vito Latora.
\newblock Structural measures for multiplex networks.
\newblock {\em Physical Review E}, 89(3):032804, 2014.

\bibitem{Berlingerio2012}
Michele Berlingerio, Michele Coscia, Fosca Giannotti, Anna Monreale, and Dino
  Pedreschi.
\newblock {Multidimensional networks: foundations of structural analysis}.
\newblock {\em World Wide Web}, 16(5-6):567--593.

\bibitem{Buchheim_crossingsand}
Christoph Buchheim, Markus Chimani, Carsten Gutwenger, Michael Jünger, and
  Petra Mutzel.
\newblock {\em Crossings and Planarization}.
\newblock CRC Press, 2013.

\bibitem{Cardillo2013}
Alessio Cardillo, Jes{\'{u}}s G{\'{o}}mez-Garde{\~{n}}es, Massimiliano Zanin,
  Miguel Romance, David Papo, Francisco del Pozo, and Stefano Boccaletti.
\newblock {Emergence of network features from multiplexity.}
\newblock {\em Scientific reports}, 3:1344, 2013.

\bibitem{Chen21032006}
Beth~L. Chen, David~H. Hall, and Dmitri~B. Chklovskii.
\newblock Wiring optimization can relate neuronal structure and function.
\newblock {\em Proceedings of the National Academy of Sciences of the United
  States of America}, 103(12):4723--4728, 2006.

\bibitem{de2014muxviz}
Manlio {De Domenico}, Mason~A. Porter, and Alex Arenas.
\newblock {MuxViz: a tool for multilayer analysis and visualization of
  networks}.
\newblock {\em Journal of Complex Networks}, pages 159--176, 2014.

\bibitem{Diaz:2002:SGL:568522.568523}
Josep D\'{\i}az, Jordi Petit, and Maria Serna.
\newblock A survey of graph layout problems.
\newblock {\em ACM Comput. Surv.}, 34(3):313--356, September 2002.

\bibitem{Dickison2016}
Mark~E. Dickison, Matteo Magnani, and Luca Rossi.
\newblock {\em {Multilayer Social Networks}}.
\newblock Cambridge University Press, 2016.

\bibitem{Dwyer}
Tim Dwyer, Kim Marriott, Falk Schreiber, Peter Stuckey, Michael Woodward, and
  Michael Wybrow.
\newblock {Exploration of networks using overview+detail with constraint-based
  cooperative layout}.
\newblock {\em IEEE transactions on visualization and computer graphics},
  14(6):1293--300, January 2008.

\bibitem{Fruchterman:1991:GDF:137556.137557}
Thomas M.~J. Fruchterman and Edward~M. Reingold.
\newblock Graph drawing by force-directed placement.
\newblock {\em Softw. Pract. Exper.}, 21(11):1129--1164, November 1991.

\bibitem{Huang20072821}
Xiaodi Huang, Wei Lai, A.S.M. Sajeev, and Junbin Gao.
\newblock A new algorithm for removing node overlapping in graph visualization.
\newblock {\em Information Sciences}, 177(14):2821 -- 2844, 2007.

\bibitem{10.2307/2785979}
Herbert~Menzel James~Coleman, Elihu~Katz.
\newblock The diffusion of an innovation among physicians.
\newblock {\em Sociometry}, 20(4):253--270, 1957.

\bibitem{Moody2005}
Skye~Benderâ€deMoll James~Moody, Daniel~McFarland.
\newblock Dynamic network visualization.
\newblock {\em American Journal of Sociology}, 110(4):1206--1241, 2005.

\bibitem{Letters1989}
Tomihisa~Kamada and Satoru~Kawai.
\newblock An algorithm for drawing general undirected graphs.
\newblock {\em Inf. Process. Lett.}, 31(1):7--15, April 1989.

\bibitem{Kivela2014}
Mikko Kivel{\"{a}}, Alexandre Arenas, Marc Barthelemy, James~P. Gleeson, Yamir
  Moreno, and Mason~A. Porter.
\newblock {Multilayer Networks}.
\newblock {\em Journal of Complex Networks}, 2(3):203--271, sep 2014.

\bibitem{1173159}
Yehuda Koren, L.~Carmel, and D.~Harel.
\newblock Ace: a fast multiscale eigenvectors computation for drawing huge
  graphs.
\newblock In {\em Information Visualization, 2002. INFOVIS 2002. IEEE Symposium
  on}, pages 137--144, 2002.

\bibitem{Kumar:2009:VCD:1529282.1529685}
Pushpa Kumar and Kang Zhang.
\newblock Visualization of clustered directed acyclic graphs with node
  interleaving.
\newblock In {\em Proceedings of the 2009 ACM Symposium on Applied Computing},
  SAC '09, pages 1800--1805, New York, NY, USA, 2009. ACM.

\bibitem{Kurant2006}
Maciej Kurant and Patrick Thiran.
\newblock {Layered Complex Networks}.
\newblock {\em Physical Review Letters}, 96(13):138701, apr 2006.

\bibitem{some}
Ding Ma.
\newblock {Visualization of social media data: mapping changing social netwroks
  }.
\newblock Master's thesis, the Faculty of Geo-Information Science and Earth
  Observation of the University of Twent, 2012.

\bibitem{DBLP:conf/asonam/MagnaniR11}
Matteo Magnani and Luca Rossi.
\newblock {The ML-Model for Multi-layer Social Networks}.
\newblock In {\em Proceedings of the International conference on Social Network Analysis and
  Mining (ASONAM)}, pages 5--12. IEEE Computer Society, 2011.

\bibitem{Padgett2006}
John~F Padgett and Paul~D McLean.
\newblock {Organizational Invention and Elite Transformation: The Birth of
  Partnership Systems in Renaissance Florence}.
\newblock {\em American Journal of Sociology}, 111(5):pp. 1463--1568, 2006.

\bibitem{Redondo2015}
Denis Redondo, Arnaud Sallaberry, Dino Ienco, Faraz Zaidi, and Pascal Poncelet.
\newblock {Layer-Centered Approach for Multigraphs Visualization}.
\newblock In {\em Proceedings of the International Conference on Information Visualisation (iV)},
  pages 50--55, 2015.

\bibitem{10.1111:cgf.12644}
Benjamin~Renoust, Guy~Melançon, and Tamara~Munzner.
\newblock Detangler: Visual analytics for multiplex networks.
\newblock {\em Computer Graphics Forum}, 34(3):321--330, 2015.

\bibitem{Roberts2011}
Nancy Roberts and Sean~F. Everton.
\newblock {Roberts and Everton Terrorist Data: Noordin Top Terrorist Network
  (Subset)}, 2011.

\bibitem{2015arXiv150101666M}
Luca Rossi and Matteo Magnani.
\newblock Towards effective visual analytics on multiplex and multilayer
  networks.
\newblock {\em Chaos, Solitons \& Fractals}, 72:68 -- 76, 2015.

\bibitem{salehi2015spreading}
Mostafa Salehi, Rajesh Sharma, Moreno Marzolla, Matteo Magnani, Payam Siyari,
  and Danilo Montesi.
\newblock Spreading processes in multilayer networks.
\newblock {\em Network Science and Engineering, IEEE Transactions on},
  2(2):65--83, 2015.

\bibitem{schaefer2013graph}
Marcus Schaefer.
\newblock The graph crossing number and its variants: a survey.
\newblock {\em The electronic journal of combinatorics}, 1000:DS21--May, 2013.

\bibitem{shabbeer2010optimal}
A~Shabbeer, C~Ozcaglar, M~Gonzalez, and KP~Bennett.
\newblock Optimal embedding of heterogeneous graph data with edge crossing
  constraints.
\newblock In {\em Presented at NIPS Workshop on Challenges of Data
  Visualization}, page~1, 2010.

\bibitem{Sole-Ribalta:2014:CRM:2615569.2615687}
Albert Sol{\'e}-Ribalta, Manlio De~Domenico, Sergio G\'{o}mez, and Alex Arenas.
\newblock Centrality rankings in multiplex networks.
\newblock In {\em Proceedings of the 2014 ACM Conference on Web Science},
  WebSci '14, pages 149--155, New York, NY, USA, 2014. ACM.

\bibitem{Landesberger2011}
Tatina~von Landesberger, Arjan~Kuijper, Tobias~Schreck, Jorn~Kohlhammer, Jarke van Wijk,
  Jean-Daniel. Fekete, and Dieter Fellner.
\newblock Visual analysis of large graphs: State-of-the-art and future research
  challenges.
\newblock {\em Computer Graphics Forum}, 30(6):1719--1749, 2011.

\bibitem{Wasserman1994}
Stanley Wasserman and Katherine Faust.
\newblock {\em {Social Network Analysis: Methods and Applications}}, volume~8
  of {\em Structural analysis in the social sciences, 8}.
\newblock Cambridge University Press, 1994.

\bibitem{Erten2004}
Cesim Erten, Stephen~G. Kobourov, Vu~Le, and Armand Navabi.
\newblock {\em Simultaneous Graph Drawing: Layout Algorithms and Visualization
  Schemes}, pages 437--449.
\newblock Springer Berlin Heidelberg, Berlin, Heidelberg, 2004.

\bibitem{GAJER20043}
Pawel Gajer, Michael~T. Goodrich, and Stephen~G. Kobourov.
\newblock A multi-dimensional approach to force-directed layouts of large
  graphs.
\newblock {\em Computational Geometry}, 29(1):3 -- 18, 2004.

\bibitem{peng}
Wu Peng, Li SiKun.
Social network analysis layout algorithm under ontology model.
\newblock {\em SOFTWARE}, 6(7):3 -- 18, JULY 2011.

\bibitem{Gan-2013}
Na; Ma Yao; Lu~Hongwei Gan, Zaobin;~Li.
\newblock [ieee 2013 10th web information system and application conference
  (wisa) - yangzhou, china (2013.11.10-2013.11.15)] 2013 10th web information
  system and application conference - trust network visualization based on
  force-directed layout.
\newblock 2013.

\bibitem{Baur2005}
Michael Baur, Ulrik Brandes, Marco Gaertler, and Dorothea Wagner.
\newblock {\em Drawing the AS Graph in 2.5 Dimensions}, pages 43--48.
\newblock Springer Berlin Heidelberg, Berlin, Heidelberg, 2005.
\end{thebibliography}
\end{document}